\documentstyle{article}
\begin{document}
\title{Induced Gravity Model \\
Based on External Impinging Neutrinos: \\
Calculation of G in Terms of \\
Collision Phenomena and Inferences to \\
Inertial Mass and Atomic Quantization}

\author{W. G. Stanley and G. C. Vezzoli, \\
Institute for the Basic Sciences, \\
51 Park Drive, Suite 22, Boston, MA 02215, USA}

\maketitle

\begin{abstract}

Herein, we present a particle-based mechanism and mathematical formulation of gravity, focusing on the neutrino as the gravity-inducing particle.
The mechanism is based on the primacy of momentum conservation and postulates an omni-directional distribution throughout the universe of fast-moving small particles of finite mass that have a low probability of colliding with nucleons.
The measured acceleration between two neighboring mass bodies results from an alteration of this distribution caused by nucleons of each body interacting with some of those particles.
Based on findings convincingly establishing that the neutrino has mass, we evaluate the various neutrinos as external particle candidates.
We show that for mass quantities up to several times that of the sun the mathematical form of the time rate of momentum transfer to each body is proportional to the product of the two body masses because of the probability nature of any collision process, and inversely proportional to the square of the distance between them because of the mathematical properties of an altered particle flux.
A derived expression involving the neutrino momentum flux, the neutrino collision cross section with nucleons, and the nucleon mass replaces the constant G from the classical gravitational model.
The neutrino momentum flux that is required to account for gravity is so large as to cause us herein to re-evaluate conventional notions in kinematics and the cause of inertial properties and to examine neutrino-nucleon collisions as a possible source of electromagnetic standing waves essential to establish electron shell states.
This line of reasoning indicates that in a much more massive body that is accreting additional mass, a coulombic collapse to a black hole will ensue when external neutrinos lose the ability to penetrate in sufficient numbers to the central region.

\end{abstract}

\section{Introduction}

The gravitational attraction between mass bodies that tends to accelerate two such bodies toward each other has been thought to be caused by some intrinsic property of matter to either produce an attractive central force field (herein referred to as the ``Newtonian field view'') or to eject a gravity-bearing particle (herein referred to as the ``graviton view'' [1]) which interacts with the matter of the neighboring body.
Recent advances, such as at CERN, in the areas of the fundamental particles of leptons and quarks have indicated the very fundamental nature of the momentum quantity at the expense of the concept of ``\emph{force} fields'' in the explanation of electro-nuclear forces [2].
However, these advances have not reported identification of a fundamental graviton and associated mechanism giving rise to gravity, or of a particle at the fundamental root of a unified field theory.

The inability of the aforementioned efforts to provide an accepted explanation of what gravity actually is (as contrasted to what gravity actually does) has led us to favor and critically examine a theory of gravity based on a largely uniform and omni-directional external particle distribution flooding the universe and weakly interacting with every mass body that it encounters.
(This is referred to herein as the ``external particle view''.)
Through collisions, each mass body acts as a sink that alters the external particle distribution in its vicinity and causes a small flux imbalance that is directed toward that body and has a magnitude that decreases with distance from that body.
The net transfer of incident (impinging) momentum from this altered external particle distribution to neighboring mass bodies accelerates the bodies toward each other; this theory then provides a ``push'' rather than a ``pull'' basis for gravity.

Considering the potential of the external omnidirectional particle view to explain gravity and contemplating the large particle density requirement that results has led us to attempt to explain the ill-understood inertial property of mass through a similar mechanistic description involving the same external particles.

Realizing that an omnidirectional particle viewpoint has been the subject of earlier qualitative inquiry, we contribute herein an independent rigorous mathematical exposition beginning from first principles.

\section{The Mathematical Formulation}

\subsection{Representation of the External Particle Distribution}

Statistical properties of the external particle (denoted by \( \varepsilon \)) can be described by a distribution function \( f_{\varepsilon}(x,v) \) in position and velocity space, which yields the external particle number density
\begin{equation} n_{\varepsilon}(x) \ \equiv \ \int f_{\varepsilon}(x,v) \ d^{3}v \end{equation}
and net external particle number flux (net number of external particles crossing unit area per unit time)
\begin{equation} n_{\varepsilon}(x) \langle {\bf v}_{\varepsilon}(x) \rangle \ \equiv \ \int f_{\varepsilon}(x,v) \ {\bf v} \ d^{3}v. \end{equation}
Unless otherwise explicitly stated, flux used alone means number flux throughout this paper.

For simplicity, we assume that far from any significant gravitational mass bodies the external particle background distribution over very large distances is uniform in \( {\bf x} \) and omni-directional and can be represented by the idealized mathematical form:
\begin{equation} f_{\varepsilon}(x,v) \ \Rightarrow \ F_{{\varepsilon}_{0}} \ \delta (v_{r}-v_{{\varepsilon}_{0}}). \end{equation}
This leads to a spatially uniform external particle background number density
\begin{equation} n_{\varepsilon}(x) \ \Rightarrow \ n_{{\varepsilon}_{0}} \ = \ \int F_{{\varepsilon}_{0}} \ \delta (v_{r}-v_{{\varepsilon}_{0}}) \ \cos \theta \ d\theta \ d\phi \ v_{r}^{2} \ dv_{r} \ = \ 4 \ \pi \ F_{{\varepsilon}_{0}} \ v_{{\varepsilon}_{0}}^{2}, \end{equation}
where \( v_{r} \cos \theta d\theta, v_{r} d\phi, \) and \( dv_{r} \) are the differential volume elements of velocity space.
The corresponding net external particle background flux is
\begin{equation} n_{{\varepsilon}_{0}} \langle {\bf v}_{\varepsilon}(x) \rangle \ = \ 0. \end{equation}

\subsection{Effects of a Mass Body on the External Particle \newline Distribution}

\subsubsection{Small Idealized Mass Body}

We consider the presence of a small idealized mass body \( M'_{s} \) positioned at \( {\bf x'} \) and its effect on the external particle distribution in near-by regions.
The external particles interact with the mass particles of \( M'_{s} \) as they pass through it to the extent that the interactions have a finite collision cross section.
The types of collisions considered are capture and elastic scattering of external particles by the mass particles.
We assume: small spherical volume; uniform mass particle distribution \( n_{m'}(x') = n_{m'} \) within that volume; total mass particle content sufficiently small that the resulting small total collision probabilities permit \( f_{\varepsilon}(x',v) \) to be treated throughout the volume as equal to the undiminished external particle background distribution \( F_{{\varepsilon}_{0}} \ \delta (v_{r}-v_{{\varepsilon}_{0}}) \).
Each external particle can be considered to either: pass through the volume un-obstructed; suffer a capture collision; or suffer only one elastic collision.
The number of mass particles within \( M'_{s} \) is large enough to justify a volume integral, but small enough to justify using this undiminished background distribution of external particles.

These collisions will produce in the region (\( {\bf x} \)) outside the immediate vicinity of \( M'_{s} \) an idealized altered distribution of external particles of the form:
\begin{equation} f_{\varepsilon}(x,v) \ = \ F_{{\varepsilon}_{0}} \ \delta (v_{r}-v_{{\varepsilon}_{0}}) \ - \label{eq:velocity_distribution_small} \end{equation}
\[ a_{M'_{s}} \ \delta (v_{r}-v_{{\varepsilon}_{0}}) \ \delta (\theta - \theta') \ \delta (\phi - \phi') \ / \ [ \ \cos ( \theta - \theta' ) \ v_{{\varepsilon}_{0}}^{2} \ |{\bf x}-{\bf x'}|^{2} \ ] \ - \]
\[ b_{M'_{s}} \ \delta (v_{r}-v_{{\varepsilon}_{0}}) \ \delta (\theta - \theta') \ \delta (\phi - \phi') \ / \ [ \ \cos ( \theta - \theta' ) \ v_{{\varepsilon}_{0}}^{2} \ |{\bf x}-{\bf x'}|^{2} \ ] \ + \]
\[ c_{M'_{s}} \ \delta (v_{r}-v_{{\varepsilon}_{1}}) \ \delta (\theta - \theta') \ \delta (\phi - \phi') \ / \ [ \ \cos ( \theta - \theta' ) \ v_{{\varepsilon}_{1}}^{2} \ |{\bf x}-{\bf x'}|^{2} \ ] \ \ \ , \]
where the alteration is spherically symmetric with respect to (\( {\bf x'} \))
and where \( [ \ \delta (\theta - \theta') \ \delta (\phi - \phi') \ / \ \cos ( \theta - \theta' ) \ ] \) expresses the directed nature of this distribution in velocity space, which at \( {\bf x} \) is non-zero only in the radial direction relative to \( {\bf x'} \), and which satisfies
\[ \int [ \ \delta (\theta - \theta') \ \delta (\phi - \phi')/\cos ( \theta - \theta' ) \ ] \ \ cos \theta \ d\theta \ d\phi \ = \ 1 . \]
(For notational brevity in the following, we introduce \( \delta ( v_{r}, v_{0}, \theta,  \theta', \phi, \phi' ) \) as a substitute for \( \delta (v_{r}-v_{0}) \delta (\theta - \theta') \delta (\phi - \phi')/\cos ( \theta - \theta' ) \).) 

\( a_{M'_{s}} \) accounts for removal from the distribution those incident external particles that were captured; \( b_{M'_{s}} \) accounts for removal from the distribution those incident external particles that were scattered; and \( c_{M'_{s}} \) accounts for addition of the scattered external particles (particles which move radially outward from \( {\bf x'} \) at speed \( v_{{\varepsilon}_{1}} \)) to the distribution.
\( a_{M'_{s}} \) and \( b_{M'_{s}} \) can be considered to form a deficit in the external particle distribution whereby the background external particles traveling toward \( M'_{s} \) are no longer flux-compensated by an equal number of particles (identical but of opposite velocity direction) that have already passed through \( {\bf x'} \) because of the collisions with \( M'_{s} \).
[Although the scattered external particles would have a distribution of speeds (all close to \( v_{{\varepsilon}_{0}} \) if the external particles have much less inertial mass than the mass particles with which they collide), we have also simplified the mathematical analysis by employing only one representative average speed \( v_{{\varepsilon}_{1}} \).]
All three components of the altered portion of the distribution do not arise from the external particles incident on \( M'_{s} \); rather, they are alterations superimposed \emph{only} on the portion of the external particle distribution in \emph{velocity} space that is directed \emph{away} from \( M'_{s} \).

The corresponding external particle number density at \( {\bf x} \) is:
\[ n_{\varepsilon}(x) \ = \ \int f_{\varepsilon}(x,v) \ d^{3}v \ = \ \int F_{{\varepsilon}_{0}} \ \delta (v_{r}-v_{{\varepsilon}_{0}}) \ \cos \theta \ d\theta \ d\phi \ v_{r}^{2} \ dv_{r} \ - \]
\[ \int [ \ ( \ a_{M'_{s}} \ + b_{M'_{s}} \ ) \ \delta ( v_{r}, v_{{\varepsilon}_{0}}, \theta,  \theta', \phi, \phi' )/(v_{{\varepsilon}_{0}}^{2} \ |{\bf x}-{\bf x'}|^{2}) \ ] \ \cos \theta \ d\theta \ d\phi \ v_{r}^{2} \ dv_{r} \ + \]
\[ \int [ \ c_{M'_{s}} \ \delta ( v_{r}, v_{{\varepsilon}_{1}}, \theta,  \theta', \phi, \phi' )/(v_{{\varepsilon}_{1}}^{2} \ |{\bf x}-{\bf x'}|^{2}) \ ] \ \cos \theta \ d\theta \ d\phi \ v_{r}^{2} \ dv_{r} \]
\begin{equation} = \ 4 \ \pi \ F_{{\varepsilon}_{0}} \ v_{{\varepsilon}_{0}}^{2} \ - \ [ \ a_{M'_{s}} \ + \ b_{M'_{s}} \ - \ c_{M'_{s}} \ ]/|{\bf x}-{\bf x'}|^{2}. \end{equation}
The corresponding net external particle number flux is:
\[ n_{\varepsilon}(x) \langle {\bf v}_{\varepsilon}(x) \rangle \ = \ \int f_{\varepsilon}(x,v) \ {\bf v} \ d^{3}v \ = \]
\[ - \ \int [ ( a_{M'_{s}} + b_{M'_{s}} ) \delta ( v_{r}, v_{{\varepsilon}_{0}}, \theta,  \theta', \phi, \phi' ) ({\bf x}-{\bf x'})/(v_{{\varepsilon}_{0}}^{2} |{\bf x}-{\bf x'}|^{3}) \ ] \ \cos \theta \ d\theta \ d\phi \ v_{r}^{3} \ dv_{r} \ + \]
\[ \int [ \ c_{M'_{s}} \ \delta ( v_{r}, v_{{\varepsilon}_{1}}, \theta,  \theta', \phi, \phi' ) \ ({\bf x}-{\bf x'})/(v_{{\varepsilon}_{1}}^{2} \ |{\bf x}-{\bf x'}|^{3}) \ ] \ \cos \theta \ d\theta \ d\phi \ v_{r}^{3} \ dv_{r} \]
\begin{equation} = \ [ \ - \ a_{M'_{s}} \ v_{{\varepsilon}_{0}} \ - \ b_{M'_{s}} \ v_{{\varepsilon}_{0}} \ + \ c_{M'_{s}} \ v_{{\varepsilon}_{1}} \ ] \ ({\bf x}-{\bf x'}) /|{\bf x}-{\bf x'}|^{3}. \end{equation}
It is thus seen that the net effect of this mass body on both scalar and vector external particle quantities (as a function of relative position and finite propagation velocity) is opposite to that of substituting at the same position a localized source of external particles. 

The incoming external particles that are scattered and their outgoing scattered counterparts together cause no accumulation of external particles within \( M'_{s} \).
The flux continuity equation:
\begin{equation} \partial n_{\varepsilon}(x')/\partial t \ + \ {\bf \nabla}' {\bf \cdot} [ \ n_{\varepsilon}(x') \langle {\bf v}_{\varepsilon}(x') \rangle \ ] \ = \ 0, \label{eq:continuity} \end{equation}
as applied to scattering collisions, and the conversion of the second term to a surface integral (see equations~\ref{eq:gauss1} and~\ref{eq:gauss2} below for procedural details) involving the combined flux of the \( b_{M'_{s}} \) and \( c_{M'_{s}} \) constituents dictate that if there is no accumulation of scattered external particles within \( M'_{s} \), their flux contributions external to \( M'_{s} \) sum to zero:
\[ - \ b_{M'_{s}} \ v_{{\varepsilon}_{0}} \ + \ c_{M'_{s}} \ v_{{\varepsilon}_{1}} \ = \ 0. \]
\begin{equation} c_{M'_{s}} \ = \ b_{M'_{s}} \ v_{{\varepsilon}_{0}}/v_{{\varepsilon}_{1}}. \label{eq:scatter_flux_balance} \end{equation} 

In the following we evaluate both \( a_{M'_{s}} \) and \( b_{M'_{s}} \) in terms of appropriate collision cross sections, time rates of collisions, and flux divergence.
Throughout the volume of \( M'_{s} \), the number density of mass particles was specified above as being uniform (\( n_{m'} \)).
The time rate of capture collisions per unit volume between the mass particles and the external particles is:
\begin{equation} \partial n_{\varepsilon}(x')/\partial t \ = \ \int \ f_{\varepsilon} (x', v_{r}) \ v_{r} \ \cos \theta \ d\theta \ d\phi \ v_{r}^{2} \ dv_{r} \ \sigma_{\varepsilon,m',cc} \ n_{m'}, \label{eq:collision_rate_per_vol_per_vol} \end{equation}
where \( \sigma_{\varepsilon,m',cc} \) is the cross section area for capture collisions between external particles and \( m' \) particles.
The integral on the right is a probabilistic quantity that requires averaging over many collisions, and contains a flux-like term \( f_{\varepsilon} (x', v_{r}) \ v_{r} \).

Throughout the small mass volume, \( f_{\varepsilon} (x', v_{r}) \) is approximated by the external particle background distribution \( F_{{\varepsilon}_{0}} \ \delta (v_{r}-v_{{\varepsilon}_{0}}) \).
Integrating over the entire mass body \( M'_{s} \) yields the time rate of capture collisions per unit volume of external particles:
\[ \int \ [ \partial n_{\varepsilon}(x')/\partial t ] \ d^{3}x' \ \simeq \]
\[ \int \ F_{{\varepsilon}_{0}} \ \delta (v_{r}-v_{{\varepsilon}_{0}}) \ v_{r} \ \cos \theta \ d\theta \ d\phi \ v_{r}^{2} \ dv_{r} \ \sigma_{\varepsilon,m',cc} \ \int \ n_{m'} \ d^{3}x' \ = \]
\begin{equation} 4 \ \pi \ F_{{\varepsilon}_{0}} \ v_{{\varepsilon}_{0}}^{3} \ \sigma_{\varepsilon,m',cc} \ N_{M'_{s}} \ = \ n_{{\varepsilon}_{0}} \ v_{{\varepsilon}_{0}} \ \sigma_{\varepsilon,m',cc} \ N_{M'_{s}}, \label{eq:collision_rate} \end{equation}
where \( N_{M'_{s}} \) is the total number of mass particles in \( M'_{s} \). 

In accordance with the flux continuity equation, this time rate of capture collisions has to be compensated by a net inward external particle flux.
Integrating equation~\ref{eq:continuity} as applied to capture collisions over the entire region of \( M'_{s} \) and substituting the results of equation~\ref{eq:collision_rate} yields
\begin{equation} \int {\bf \nabla}' {\bf \cdot} [ \ n_{\varepsilon}(x') \langle {\bf v}_{\varepsilon}(x') \rangle \ ] \ d^{3}x' \ = \ - \ n_{{\varepsilon}_{0}} \ v_{{\varepsilon}_{0}} \ \sigma_{\varepsilon,m',cc} \ \ N_{M'_{s}}. \label{eq:gauss1} \end{equation}

Converting the volume integral of external particle flux divergence on the left to a surface integral over a spherical surface outside the immediate vicinity of \( M'_{s} \) and evaluating that term yields
\[ \oint d{\bf a'} {\bf \cdot} [ \ n_{\varepsilon}(x') \langle {\bf v}_{\varepsilon}(x') \rangle \ ] \ = \]
\[ - \ a_{M'_{s}} \ v_{{\varepsilon}_{0}} \ \oint [ \ |{\bf x}-{\bf x'}|^{2} \ ({\bf x}-{\bf x'})/|{\bf x}-{\bf x'}| \ ] {\bf \cdot} [ \ ({\bf x}-{\bf x'})/|{\bf x}-{\bf x'}|^{3} \ ] \ \cos \vartheta \ d\vartheta \ d\varphi \]
\begin{equation} = \ - \ 4 \ \pi \ a_{M'_{s}} \ v_{{\varepsilon}_{0}}, \label{eq:gauss2} \end{equation}
where only the \( a_{M'_{s}} \) constituent of the flux survived because the \( b_{M'_{s}} \) and \( c_{M'_{s}} \) constituents of flux sum to zero everywhere outside \( M'_{s} \).

So,
\[ 4 \ \pi \ a_{M'_{s}} \ v_{{\varepsilon}_{0}} \ = \ n_{{\varepsilon}_{0}} \ v_{{\varepsilon}_{0}} \ \sigma_{\varepsilon,m',cc} \ N_{M'_{s}} \]
and
\begin{equation} a_{M'_{s}} \ = \ n_{{\varepsilon}_{0}} \ \sigma_{\varepsilon,m',cc} \ N_{M'_{s}}/(4 \ \pi) . \end{equation}
To the extent that there is capture, the net flux is everywhere directed radially inward relative to the position of \( M'_{s} \).

Before proceeding further, it is necessary to introduce and define two additional cross section quantities: the scattering cross section for total number flux transfer
\begin{equation} \sigma_{sc,tnf} \equiv \label{eq:scattering_cross_section_for_number_flux_transfer} \end{equation}
\[ \{ \int 2 \pi s \ [v_{0} - v'(\vartheta(s)) \cos \vartheta(s)] \ ds \} \ / \ v_{0}; \] 
and the scattering cross section for total momentum flux transfer
\begin{equation} \sigma_{sc,tpf} \equiv \label{eq:scattering_cross_section_for_momentum_flux_transfer} \end{equation}
\[ \{ \int 2 \pi s \ [v_{0} m_{0} v_{0} - v'(\vartheta(s)) m'(\vartheta(s)) v'(\vartheta(s)) \cos \vartheta(s)] \ ds \} \ / \ ( v_{0} m_{0} v_{0} ). \]
These both express probabilities of ``effective'' scattering collisions and refer to the normalized target area that will exactly remove (by deflection) the original vector component of the designated flux quantity.
They are calculated so as to be similar in meaning to the capture collision cross section which, by definition, completely absorbs (at least for an instant) the particle and any momentum or energy that it is carrying.
They are defined herein for the convenience of concisely expressing particle number, momentum, and energy transfer parameters, and they are effective because the resulting cross sections are very small compared to the profile area of the nucleon. 

As with the time rate of capture collisions, the volume integral of the time rate of external particle momentum flux transfer from scattering and the surface integral of the appropriate deficit portion (\( b_{M'_{s}} \)) of the flux that results from this scattering yield
\begin{equation} b_{M'_{s}} \ = \ n_{{\varepsilon}_{0}} \ \sigma_{\varepsilon,m',sc,tnf} \ N_{M'_{s}}/(4 \ \pi). \end{equation}
The scattered speeds are equal to or lower than the speeds of the incident external particles (\( v_{{\varepsilon}_{1}} \le v_{{\varepsilon}_{0}} \)).

Fully expanded, the distribution function is
\begin{equation} f_{\varepsilon}(x,v) \ = \ F_{{\varepsilon}_{0}} \ \delta (v_{r}-v_{{\varepsilon}_{0}}) \ - \label{eq:velocity_distribution} \end{equation}
\[ n_{{\varepsilon}_{0}} \ ( \sigma_{\varepsilon,m',cc} + \sigma_{\varepsilon,m',sc,tnf} ) \ N_{M'} \ \delta ( v_{r}, v_{{\varepsilon}_{0}}, \theta,  \theta', \phi, \phi' )/(4 \ \pi \ v_{{\varepsilon}_{0}}^{2} \ |{\bf x}-{\bf x'}|^{2}) \ + \]
\[ n_{{\varepsilon}_{0}} \ \sigma_{\varepsilon,m',sc,tnf} \ N_{M'} \ \delta ( v_{r}, v_{{\varepsilon}_{1}}, \theta,  \theta', \phi, \phi' ) \ v_{{\varepsilon}_{0}}/(4 \ \pi \ v_{{\varepsilon}_{1}}^{3} \ |{\bf x}-{\bf x'}|^{2}) . \]The corresponding number density is
\begin{equation} n_{\varepsilon}(x) \ = \ 4 \ \pi \ F_{{\varepsilon}_{0}} \ v_{{\varepsilon}_{0}}^{2} \ - \end{equation}
\[ n_{{\varepsilon}_{0}} \ ( \sigma_{\varepsilon,m',cc} + \sigma_{\varepsilon,m',sc,tnf} ) \ N_{M'}/(4 \ \pi \ |{\bf x}-{\bf x'}|^{2}) \ + \]
\[ n_{{\varepsilon}_{0}} \ \sigma_{\varepsilon,m',sc,tnf} \ N_{M'} \ v_{{\varepsilon}_{0}} /(4 \ \pi \ v_{{\varepsilon}_{1}} \ |{\bf x}-{\bf x'}|^{2}), \]
and the net external particle number flux is
\begin{equation} n_{\varepsilon}(x) \langle {\bf v}_{\varepsilon}(x) \rangle \ = \ - \ n_{{\varepsilon}_{0}} \ v_{{\varepsilon}_{0}} \ \sigma_{\varepsilon,m',cc} \ N_{M'} \ ({\bf x}-{\bf x'})/(4 \ \pi \ |{\bf x}-{\bf x'}|^{3}) . \end{equation}

\subsubsection{Finite Mass Body}

In attempting to apply the above approach to the larger mass bodies of interest that occur in nature, complicating issues arise such as: spatially variable mass particle density within the body, attenuation of the external particle distribution within the body caused by collisions with internal mass particles, possibly large numbers of scattering collisions suffered by each external particle, and continuous distributions of both post-scattering speeds and angles.
Some of these difficulties must be simplified in order to determine how, using this external particle approach, two mass bodies affect each other's time rate of change of momentum and how the results differ in principle from any other accepted approach.
We will carry complex integral expressions (unreduced) or expand them and retain first and second-order expansion terms when they account for a significant portion of the important phenomena and predicted differences.
The approximation procedures and corresponding results obtained in this section will likely apply only to planets and, possibly, to small stars.
The approximations will not be able to be applied to hugely massive bodies, but the general tendency of the total mass particle content of a body to be limited (as contrasted with the Newtonian view) in its ability to gravitationally influence other mass bodies will already be evident in the finite mass body expressions that are calculated.

Assuming low collision probabilities between the external particles and mass particles there is no zeroth-order effect on a finite mass body (\( M'_{f} \)) because almost all external particles pass unaffected through it.
The first-order effect on a body causes compression [3].
Also, given that some external particles are removed or slowed by collisions with that first body, a first-order directional effect is caused in regions external to the first body, where there is a \emph{net deficit} in the external particle distribution because of removal of (or alteration to) those external particles that would otherwise have passed unaffected through the region occupied by the first body.
To first order, this distribution deficit has no effect on a second nearby finite body.
This is because the small number of background external particles incident at the second body that are not now completely compensated in all respects by the external particles that have been captured (or slowed by scattering) by the first body are then unlikely to collide with the second body.
The effect of such uncompensated incident external particles on the second body is a second-order effect, is directional (toward the first body), and imparts a corresponding momentum change to the second body.

The general form of equation~\ref{eq:collision_rate_per_vol_per_vol} that must instead be computed for finite bodies is
\begin{equation} \int \ \int \ f_{\varepsilon} (x', v_{r}) \ v_{r} \ \cos \theta \ d\theta \ d\phi \ v_{r}^{2} \ dv_{r} \ \sigma_{\varepsilon,m',cc} \ n_{m'}(x') \ d^{3}x'. \end{equation}
In this external particle approach to gravity, the integral over the mass body \( M'_{f} \) is greatly complicated by the attenuation of the external particle distribution, whereas the Newtonian field view attributes no such attenuation within \( M'_{f} \) (or anywhere else) to the gravitational field emanating from each differential mass volume.

Henceforth, we consider only bodies whose mass distribution is spherically symmetric [4]:
\[ n_{m'}(x') \ \Rightarrow \ n_{m'}(r'), \]
which accurately characterizes most large bodies.
This requirement greatly simplifies the integration over \( M'_{f} \) and allows us to again assume a spherically symmetric functional form similar to that in equation~\ref{eq:velocity_distribution_small} for the idealized external particle distribution function \emph{outside} the immediate vicinity of \( M'_{f} \):
\begin{equation} f_{\varepsilon}(x,v) \ = \ F_{{\varepsilon}_{0}} \ \delta (v_{r}-v_{{\varepsilon}_{0}}) \ - \label{eq:velocity_distribution_finite} \end{equation}
\[ ( \ a_{M'_{f}} \ + \ b_{M'_{f}} \ ) \ \delta (v_{r}, v_{{\varepsilon}_{0}}, \theta, \theta', \phi, \phi')/(v_{{\varepsilon}_{0}}^{2} \ |{\bf x}-{\bf x'}|^{2}) \ + \]
\[ c_{M'_{f}} \ \delta (v_{r}, v_{{\varepsilon}_{1}}, \theta, \theta', \phi, \phi')/(v_{{\varepsilon}_{1}}^{2} \ |{\bf x}-{\bf x'}|^{2}) , \]where \( a_{M'_{f}}, b_{M'_{f}}, c_{M'_{f}} \) play roles similar to \( a_{M'_{s}}, b_{M'_{s}}, c_{M'_{s}} \) except that they apply to much larger \( M'_{f} \) bodies rather than just to small idealized \( M'_{s} \) bodies.

We have chosen to continue representing the scattered distribution of speeds as instead a single representative speed, and we continue to assume that at most one scattering collision within \( M'_{f} \) will occur per external particle.
(If many were to instead occur, such multiply-scattered external particles make contributions that in the limit approximate the behavior of the captured external particles.)
The relationship
\begin{equation} c_{M'_{f}} \ = \ b_{M'_{f}} \ v_{{\varepsilon}_{0}}/v_{{\varepsilon}_{1}} \end{equation}
follows because the same reasoning concerning flux continuity still applies here as was used in developing equation~\ref{eq:scatter_flux_balance}.

\( a_{M'_{f}} \) and \( b_{M'_{f}} \) must again be evaluated in terms of appropriate collision cross sections, time rates of collisions, and flux divergence.
The volume integral of the time rate of capture collisions (similar to equation~\ref{eq:collision_rate}) can now be written:
\[ \int \ [ \partial n_{\varepsilon}(x')/\partial t ] \ d^{3}x' \ = \]
\begin{equation} \int \ \int \ f_{\varepsilon} (x', v_{r}) \ v_{r} \ \cos \theta \ d\theta \ d\phi \ v_{r}^{2} \ dv_{r} \ \sigma_{\varepsilon,m',cc} \ n_{m'}(r') \ d^{3}x'. \label{eq:attenuated_collision_rate} \end{equation}
The attenuation of flux-like \( f_{\varepsilon} (x', v_{r}) \ v_{r} \) from its effective value \( F_{{\varepsilon}_{0}} \ \delta (v_{r}-v_{{\varepsilon}_{0}}) \ v_{r} \) for incident external particles just before beginning to penetrate \( M'_{f} \) is treated in a probabilistic way by multiplying \( F_{{\varepsilon}_{0}} \) by an angular attenuation factor:
\[ e^{-\sigma_{\varepsilon,m',cc} \ \lambda [{\bf x'},n_{m'}(x'),\hat{\bf v}_{r}]}, \]
where \( \sigma_{\varepsilon,m',cc} \ \lambda [{\bf x'},n_{m'}(x'),\hat{\bf v}_{r}] \) represents the ever-increasing probability (along the path of travel through \( M'_{f} \) to the \( {\bf x'} \) position) that incident external particles in a differential volume \( v_{r}\cos \theta d\theta v_{r}d\phi dv_{r} \) will be removed from the external particle flux before reaching \( {\bf x'} \).
\[ \lambda [{\bf x'},n_{m'}(x'),\hat{\bf v}_{r}] \ = \ \int_{R'}^{{\bf x'}} \ n_{m''}(x'') \ dl(x'',\hat{\bf v}_{r}). \]
\( dl(x'',\hat{\bf v}_{r}) \) denotes the differential path through \( M'_{f} \) (and before arriving at \( {\bf x'} \)) for particles whose velocities lie in the \( \hat{\bf v}_{r} \) direction.
The important contribution to this attenuation is from capture collisions because scattering collisions cause little or no net change to the external particle flux.
It is for this reason that only the capture cross section appears in the attenuating exponent term that is integrated along the path of travel.

The right side of Equation~\ref{eq:attenuated_collision_rate}, incorporating this attenuation term, is now written:
\[ \int F_{{\varepsilon}_{0}} \delta (v_{r}-v_{{\varepsilon}_{0}}) v_{r} \cos \theta d\theta d\phi v_{r}^{2} dv_{r} \sigma_{\varepsilon,m',cc} \int e^{-\sigma_{\varepsilon,m',cc} \ \lambda [{\bf x'},n_{m'}(r'),\hat{\bf v}_{r}]} \ n_{m'}(r') d^{3}x'. \]

For example, if \( n_{m'}(x') \) is uniform throughout \( M'_{f} \),
\[ \lambda [{\bf x'},n_{m'}(x'),\hat{\bf v}_{r}] \ = \ n_{m'}R' \ [ \sqrt{1-(r'^{2}/R'^{2})\cos^{2}\vartheta'}-(r'/R')\sin \vartheta'], \]
where \( \vartheta' \) is the angle to the differential volume \( r'\cos \vartheta' d\vartheta' r' d\varphi' dr'\) in the spatial coordinate system \( {\bf x'} \) oriented so that its positive \( z' \) axis lies opposite to the direction to the differential volume in \( v_{r} \) space (\( \hat{\bf v}_{r} {\bf \cdot} \hat{\bf z} = -1 \)).
In this case, the volume integral is
\[ \int \ F_{{\varepsilon}_{0}} \ \delta (v_{r}-v_{{\varepsilon}_{0}}) \ v_{r} \ \cos \theta \ d\theta \ d\phi \ v_{r}^{2} \ dv_{r} \ \sigma_{\varepsilon,m',cc} \ \int \ \cos \vartheta' \ d\vartheta' \ d\varphi' \ r'^{2} \ dr' \ \cdot \]
\[ e^{ - \sigma_{\varepsilon,m',cc} \ n_{m'} \ R' \ [ \sqrt{1-(r'^{2}/R'^{2})\cos^{2}\vartheta'} \ - \ (r'/R')\sin \vartheta' ] } \ n_{m'} \]
If \( \sigma_{\varepsilon,m',cc} \ n_{m'} \ R' \ [ \sqrt{1-(r'^{2}/R'^{2})\cos^{2}\vartheta'} \ - \ (r'/R')\sin \vartheta' ] \) is relatively small, the exponent term can be replaced by the first few terms of the expansion:
\[ 1 - \sigma_{\varepsilon,m',cc} \ n_{m'} \ R' \ [ \sqrt{1-(r'^{2}/R'^{2})\cos^{2}\vartheta'} \ - \ (r'/R')\sin \vartheta' ] + \ldots \]
Retaining only the first two terms,
\[ \int \cos \vartheta' d\vartheta' d\varphi' r'^{2} dr' \{ 1 - \sigma_{\varepsilon,m',cc} n_{m'} R' [ \sqrt{1-(r'^{2}/R'^{2})\cos^{2}\vartheta'} - (r'/R')\sin \vartheta' ] \} \ n_{m'} \]
\begin{equation} = \ ( 1 \ - \ 3 \ \sigma_{\varepsilon,m',cc} \ n_{m'} \ R'/4 \ ) \ N_{M'_{f}}. \label{eq:number_deficit_example} \end{equation}
This will lead to a gravitational under-representation of the total number of mass particles in \( M'_{f} \) by the collision probability \( ( 3 \ \sigma_{\varepsilon,m',cc} \ n_{m'} \ R'/4 ) \) along the expectation path length \( 3 R'/4 \) of travel for the external particles.

Equation~\ref{eq:attenuated_collision_rate}, using the more general mass particle distribution, can be re-expressed as:
\begin{equation} \int \ [ \partial n_{\varepsilon}(x')/\partial t ] \ d^{3}x' \ = \ n_{{\varepsilon}_{0}} \ v_{{\varepsilon}_{0}} \ \sigma_{\varepsilon,m',cc} \ Q'_{M'_{f}} \ N_{M'_{f}}, \end{equation}
where \( Q'_{M'_{f}} \) is a normalized integral expression:
\[ Q'_{M'_{f}} \ = \ (4 \pi N_{M'_{f}})^{-1} \ \cdot \]
\begin{equation} \int \delta (v_{r}-v_{{\varepsilon}_{0}}) \cos \theta d\theta d\phi dv_{r} \int e^{-\sigma_{\varepsilon,m',cc} \lambda [{\bf x'},n_{m'}(r'),\hat{\bf v}_{r}]} \ n_{m'}(r') \ d^{3}x' \label{eq:attenuated_number} \end{equation}
to be evaluated once \( n_{m'}(r') \) is specified.
\( Q'_{M'_{f}} \) is generally less than \( 1 \), but in the limit of very small \( \sigma_{\varepsilon,m',cc} \ \lambda [{\bf x'},n_{m'}(r'),\hat{\bf v}_{r}] \), \( Q'_{M'_{f}} \) is \( 1 \).

The evaluation of \( a_{M'_{f}} \) and \( b_{M'_{f}} \) then proceeds as in the preceding subsection to yield:
\begin{equation} a_{M'_{f}} \ = \ n_{{\varepsilon}_{0}} \ \sigma_{\varepsilon,m',cc} \ Q'_{M'_{f}} \ N_{M'_{f}}/(4 \ \pi) ; \label{eq:attenuated_capture_strength} \end{equation}
\begin{equation} b_{M'_{f}} \ = \ n_{{\varepsilon}_{0}} \ \sigma_{\varepsilon,m',sc,tnf} \ Q'_{M'_{f}} \ N_{M'_{f}}/(4 \ \pi). \end{equation}
It should be noted that the quantity \( Q'_{M'_{f}} \) is computed in the same way for \( b_{M'_{f}} \) as for \( a_{M'_{f}} \) because both types of collisions depend on a flux-like term \( f_{\varepsilon} (x', v_{r}) \ v_{r} \) that is attenuated mainly by capture collisions.

The external particle distribution function outside the immediate vicinity of \( M'_{f} \) can now be written:
\begin{equation} f_{\varepsilon}(x,v) \ = \ F_{{\varepsilon}_{0}} \ \delta (v_{r}-v_{{\varepsilon}_{0}}) \ - \label{eq:attenuated_velocity_distribution} \end{equation}
\[ n_{{\varepsilon}_{0}} \ ( \sigma_{\varepsilon,m',cc} + \sigma_{\varepsilon,m',sc,tnf} ) \ Q'_{M'_{f}} \ N_{M'_{f}} \ \delta ( v_{r}, v_{{\varepsilon}_{0}}, \theta,  \theta', \phi, \phi' )/(4 \ \pi \ v_{{\varepsilon}_{0}}^{2} \ |{\bf x}-{\bf x'}|^{2}) \ + \]
\[ n_{{\varepsilon}_{0}} \ \sigma_{\varepsilon,m',sc,tnf} \ Q'_{M'_{f}} \ N_{M'_{f}} \ \delta ( v_{r}, v_{{\varepsilon}_{1}}, \theta,  \theta', \phi, \phi' ) \ v_{{\varepsilon}_{0}}/(4 \ \pi \ v_{{\varepsilon}_{1}}^{3} \ |{\bf x}-{\bf x'}|^{2}) . \]
The corresponding number density is
\begin{equation} n_{\varepsilon}(x) \ = \ 4 \ \pi \ F_{{\varepsilon}_{0}} \ v_{{\varepsilon}_{0}}^{2} \ - \end{equation}
\[ n_{{\varepsilon}_{0}} \ ( \sigma_{\varepsilon,m',cc} + \sigma_{\varepsilon,m',sc,tnf} \ ) \ Q'_{M'_{f}} \ N_{M'_{f}}/(4 \ \pi \ |{\bf x}-{\bf x'}|^{2}) \ + \]
\[ n_{{\varepsilon}_{0}} \ \sigma_{\varepsilon,m',sc,tnf} \ Q'_{M'_{f}} \ N_{M'_{f}} \ v_{{\varepsilon}_{0}} /(4 \ \pi \ v_{{\varepsilon}_{1}} \ |{\bf x}-{\bf x'}|^{2}), \]
and the net external particle number flux is
\begin{equation} n_{\varepsilon}(x) \langle {\bf v}_{\varepsilon}(x) \rangle \ = \ - \ n_{{\varepsilon}_{0}} \ v_{{\varepsilon}_{0}} \ \sigma_{\varepsilon,m',cc} \ Q'_{M'_{f}} \ N_{M'_{f}} \ ({\bf x}-{\bf x'})/(4 \ \pi \ |{\bf x}-{\bf x'}|^{3}) . \end{equation}

If \( Q'_{M'_{f}} \) departs significantly from \( 1.0 \), this mechanism and the way it affects the external particle distribution are both very different in character from the gravitational field calculated in a Newtonian fashion from a spherically symmetric distribution of mass particles.
We selected an observation point \( x \) (inside or outside the body) and two source points (\( x'_{near} \) and \( x'_{opp} \)) that both lie within the body equi-distant from its center and along the line that passes through its center and through \( x \).
The region on the opposite part of the body (\( x'_{opp} \)) has a disproportionately large influence on the outward-directed portion of the distribution function \( \theta'_{out} = \pi / 2 \) at \( x \) because \( f_{\varepsilon} (x'_{opp}, \theta'_{in} (= \pi / 2)) \) on that opposite side is itself larger than \( f_{\varepsilon} (x'_{near}, \theta'_{out}) \) for that same angular portion of the distribution in velocity space.
(The \( {\bf v_{r}} \) coordinate systems have been aligned with each other at both \( x'_{near} \) and \( x'_{opp} \).)
Thus, the different regions of the body do not contribute to the distribution simply as the product of the mass particle density and the inverse square of the distance between the source and observation points. 
The \emph{opposite} side of the body captures more of the external particles that ordinarily would have traveled through the body to reach \( x \) to contribute to the outward-directed portion of the distribution at \( x \).

From the above, if \( Q'_{M'_{f}} \) differs from \( 1.0 \) by just a few percent, the actual modified distribution function in the near field region is not accurately represented by equation~\ref{eq:attenuated_velocity_distribution} and has a diminished strength over a broader angular range (\( 0 \le (\theta - \theta') < \pi / 2 \)) rather than just at \( 0 = (\theta - \theta') \).
For very large finite bodies (e.g., where \( Q'_{M'_{f}} < 0.9 \)) the use of a \( \delta ( v_{r}, v_{{\varepsilon}_{0}}, \theta,  \theta', \phi, \phi' ) \) in the expression for \( f_{\varepsilon}(x,v) \) is therefore not appropriate in the near field region because it conflicts with the actual requirement that the distribution express an attenuation effect for velocity angles other than those strictly aligned with the radial direction with respect to the center of the body.

The next sections treat finite gravitational bodies and carry the \( Q'_{M'_{f}} \) term because it is close to but not exactly equal to \( 1.0 \).

\subsection{Time Rate of Momentum Transfer to a Second Nearby Finite Mass Body}

Another finite mass body \( M \) at \( {\bf x} \) (leaving off the \( f \) subscript and assuming in this section that all mass bodies are finite or small) that can also interact with the flux of external particles will capture those external particles at the time rate of
\[ \int f_{\varepsilon}(x,v_{r}) \ v_{r} \ \cos \theta \ d\theta \ d\phi \ v_{r}^{2} \ dv_{r} \ \sigma_{\varepsilon,m,cc} \ n_{m}(x) \]
per unit volume and will scatter them at the time rate of
\[ \int f_{\varepsilon}(x,v_{r}) \ v_{r} \ \cos \theta \ d\theta \ d\phi \ v_{r}^{2} \ dv_{r} \ \sigma_{\varepsilon,m,sc,tnf} \ n_{m}(x). \]

The total time rate of momentum transfer to \( M \) by this mechanism is obtained by multiplying the constituents of \( f_{\varepsilon}(x,v_{r}) \) (from equation~\ref{eq:attenuated_velocity_distribution}) in the above two expressions by the appropriate vector momentum transfer contributions for each type of collision, summing, and integrating over the volume containing \( M \).
The first constituent of \( f_{\varepsilon}(x,v_{r}) v_{r} \):
\[ F_{{\varepsilon}_{0}} \ \delta (v_{r}-v_{{\varepsilon}_{0}}) \ v_{r} \]
contributes nothing to such a vector momentum quantity (either by capture or by elastic scattering) because it represents an omni-directional distribution that produces no net directional effect on \( M \).
The next constituent of \( f_{\varepsilon}(x,v_{r}) v_{r}\):
\[ - \ n_{{\varepsilon}_{0}} \ ( \sigma_{\varepsilon,m',cc} + \sigma_{\varepsilon,m',sc,tnf} ) \ Q'_{M'} \ N_{M'} \ \delta ( v_{r}, v_{{\varepsilon}_{0}}, \theta,  \theta', \phi, \phi' )/(4 \ \pi \ v_{{\varepsilon}_{0}}^{2} \ |{\bf x}-{\bf x'}|^{2}) \]
represents a deficit taken from the background flux of external particles by \( M' \), so one must consider the time rate of momentum transfer to \( M \) from this constituent as arising from the uncompensated background flux arriving at \( M \) from the opposite side and traveling in the \( ({\bf x'}-{\bf x}) \) direction.
These external particles of relativistic mass \( m_{{\varepsilon}_{0}} \) (\emph{not} rest mass) arrive with a speed \( v_{{\varepsilon}_{0}} \) and suffer both capture and elastic scattering collisions with the mass particles of \( M \) (presumably by mechanisms very similar to the collisions with \( M' \)).
Each capture collision transfers momentum of \( m_{{\varepsilon}_{0}} {\bf v}_{{\varepsilon}_{0}} \) to \( M \) and is characterized by an arrival speed \( v_{{\varepsilon}_{0}} \) and a capture collision cross section \( \sigma_{\varepsilon,m,cc} \).
Each scattering collision transfers momentum of average magnitude \( ( m_{{\varepsilon}_{0}} v_{{\varepsilon}_{0}} - \langle m_{{\varepsilon}_{1}} v_{{\varepsilon}_{1}} cos \vartheta \rangle ) \) along the original direction of motion to \( M \) and is characterized by an arrival speed \( v_{{\varepsilon}_{0}} \), exit speed \( v_{{\varepsilon}_{1}} \), and scattering cross section for momentum flux transfer \( \sigma_{\varepsilon,m,sc,tpf} \) (see equation~\ref{eq:scattering_cross_section_for_momentum_flux_transfer}).
The last constituent of \( f_{\varepsilon}(x,v) \):
\[ n_{{\varepsilon}_{0}} \ \sigma_{\varepsilon,m',sc,tnf} \ Q'_{M'} \ N_{M'} \ \delta ( v_{r}, v_{{\varepsilon}_{1}}, \theta, \theta', \phi, \phi' ) \ v_{{\varepsilon}_{0}}/(4 \ \pi \ v_{{\varepsilon}_{1}}^{3} \ |{\bf x}-{\bf x'}|^{2}) \]
represents a flux of altered scattered external particles emanating from \( M' \) and arriving at \( M \) traveling in the \( ({\bf x}-{\bf x'}) \) direction.
These external particles presumably have almost the same rest-mass as the original external particles but arrive with a speed \( v_{{\varepsilon}_{1}} \) and suffer both capture and elastic scattering collisions with the mass particles of \( M \) (again, presumably by mechanisms very similar to the collisions with \( M' \)).
Each capture collision transfers momentum \( m_{{\varepsilon}_{1}} {\bf v}_{{\varepsilon}_{1}} \) to \( M \) and is characterized by an arrival speed \( v_{{\varepsilon}_{1}} \) and a capture collision cross section \( \sigma_{\varepsilon,m,cc} \).
Each scattering collision transfers momentum of average magnitude \( ( m_{{\varepsilon}_{1}} v_{{\varepsilon}_{1}} - \langle m_{{\varepsilon}_{2}} v_{{\varepsilon}_{2}} cos \vartheta \rangle ) \) along the original direction of motion to \( M \) and is characterized by an arrival speed \( v_{{\varepsilon}_{1}} \), exit speed \( v_{{\varepsilon}_{2}} \), and scattering cross section for momentum flux transfer \( \sigma_{\varepsilon,m,sc,tpf} \).

\( d{\bf P}_{M}/dt \) is computed by integrating each of these six momentum transfer terms over velocity space and over the spatial volume containing the mass particles of \( M \) to yield
\begin{equation} d{\bf P}_{M}/dt \ = \ - \ n_{{\varepsilon}_{0}} \ Q'_{M'} \ N_{M'} \ \Psi \ Q_{M} \ N_{M} \ ({\bf x}-{\bf x'})/(4 \ \pi \ |{\bf x}-{\bf x'}|^{3}), \end{equation}
where \pagebreak
\begin{equation} \Psi \ \equiv \label{eq:momentum_flux_coupling_term} \end{equation}
\[ v_{{\varepsilon}_{0}} \sigma_{\varepsilon,m',cc} m_{{\varepsilon}_{0}} v_{{\varepsilon}_{0}} \sigma_{\varepsilon,m,cc}  + v_{{\varepsilon}_{0}} \sigma_{\varepsilon,m',cc} m_{{\varepsilon}_{0}} v_{{\varepsilon}_{0}} \sigma_{\varepsilon,m,sc,tpf} \ + \]
\[ v_{{\varepsilon}_{0}} \sigma_{\varepsilon,m',sc,tnf} m_{{\varepsilon}_{0}} v_{{\varepsilon}_{0}} \sigma_{\varepsilon,m,cc} + v_{{\varepsilon}_{0}} \sigma_{\varepsilon,m',sc,tnf} m_{{\varepsilon}_{0}} v_{{\varepsilon}_{0}} \sigma_{\varepsilon,m,sc,tpf} \ - \]
\[ v_{{\varepsilon}_{1}} \sigma_{\varepsilon,m',sc,tnf} m_{{\varepsilon}_{1}} v_{{\varepsilon}_{1}} \sigma_{\varepsilon,m,cc} - v_{{\varepsilon}_{1}} \sigma_{\varepsilon,m',sc,tnf} m_{{\varepsilon}_{1}} v_{{\varepsilon}_{1}} \sigma_{\varepsilon,m,sc,tpf} \]
and represents a momentum flux coupling term that describes the rate of momentum transferred to a mass particle of body \( M \) per external particle that enters body \( M' \) per second per mass particle of body \( M' \).

In the above equation for \( d{\bf P}_{M}/dt \), a similar attenuation for the external particle distribution has been computed within \( M \) in case the size of \( M \) is large enough to warrant it, and
\[ \int \delta ( v_{r}, v_{{\varepsilon}_{0}}, \theta,  \theta', \phi, \phi' ) \cos \theta d\theta d\phi dv_{r} \int e^{-\sigma_{\varepsilon,m,cc} \ \lambda [{\bf x},n_{m}(x),\hat{\bf v}_{r}]} \ n_{m}(x) \ d^{3}x \]
has been notationally replaced by \( Q_{M} N_{M} \).
Although \( Q_{M} \) is calculated differently than \( Q'_{M'} \) (because of the omni-directional nature of the incident distribution at \( M' \) and the uni-directional nature of the contributing portion of the incident distribution at \( M \)), they both equal \( 1.0 \) in the limit of small \( M' \) and \( M \) and are otherwise less than \( 1.0 \).  

As long as secondary scattering can be ignored within both bodies, \( \Psi \) represents details of the capture and scattering collisions and is a function of the interactions between individual external particles and individual mass particles and is \emph{not} a function of the size, shape, and mass particle density distribution properties of either body.
If the simplified speed distribution for external particles is replaced by a continuous distribution and if a single average scattered speed is replaced by a continuous distribution, \( \Psi \) will remain a function only of the individual interactions (albeit a more complicated function).
However, if the total collision probability between a single external particle and all mass particles within either body were to be large because of \emph{huge} mass particle content, secondary scattering collisions would play a much larger role, and \( \Psi \) would change in character and itself depend on the extensive properties of that huge body.
If this total probability of capture or scattering collisions with either body becomes large, significant mass particle content will be unrepresented in the momentum transfer expression, and it will no longer be dependent on the product of the mass particle contents.
In the limit where almost all external particles are captured by a huge gravitational body, the ability of that body to induce momentum change in a test body will be limited; that limit depends on the body's total cross section area rather than on its mass particle content.
However, the \( |{\bf x}-{\bf x'}|^{-2} \) dependence still applies at large distances for spherically symmetric mass particle distributions.
This absence of strict gravitational dependence on the total quantity of mass particles is an example of the departure of this subject particle theory from the Newtonian field view.

\subsection{Interpretation of the Measured (Cavendish) \newline Attractive Force Between Two Bodies as the External Particle - Induced Time Rate of Momentum Transfer}

The word ``mass'' has been used herein only as a label on a body and on its constituent particles that actually collide with the external particles.
It implies that these constituent particles account for a large portion of the mass content of the body, but the actual mass property of any body has not yet been directly utilized.

The attractive force between \( M \) and \( M' \) as measured by Cavendish is:
\[ - \ G \ M \ M' \ ({\bf x}-{\bf x'})/|{\bf x}-{\bf x'}|^{3}, \]
where \( G \) is the gravitational constant and \( M \) and \( M' \) (in the context of this Newtonian force) refer to the actual gravitational mass property contents of the bodies.

If an equivalence is to be established between this Newtonian gravitational force and the time rate of change of momentum associated with external particle collisions, we must assume that the number of mass particles responsible for colliding with external particles is proportional to the mass content of \( M \) and \( M' \).
We also assume that both protons and neutrons are the mass particles, consistent with a gravitational force that is proportional to atomic weight and because the external particles are likely to be neutral and therefore interact with protons and neutrons in very similar (if not absolutely identical) ways.
Stated quantitatively,
\[ M = N_{M} \ m_{np}, \ \ M' = N_{M'} \ m_{np}, \]
where \( m_{np} \) is the gravitational mass of a neutron or proton, and
\[ \sigma_{\varepsilon,m',cc} = \sigma_{\varepsilon,m,cc} = \sigma_{\varepsilon,np,cc}, \ \ \sigma_{\varepsilon,m',sc,tnf} = \sigma_{\varepsilon,np,sc,tnf}, \ \ \sigma_{\varepsilon,m,sc,tpf} = \sigma_{\varepsilon,np,sc,tpf}. \]
Other fundamental particles making up a body could also contribute exchange collisions with external particles, but their collision cross sections must be far smaller than those of the proton and neutron.
Incorporating these, we now equate the gravitational force with the above collision momentum transfer expression to obtain:
\[ - \ G \ M \ M' \ ({\bf x}-{\bf x'})/|{\bf x}-{\bf x'}|^{3} \ = \]
\[ - \ n_{{\varepsilon}_{0}} \ Q'_{M'} \ N_{M'} \ \Psi \ Q_{M} \ N_{M} \ ({\bf x}-{\bf x'})/(4 \ \pi \ |{\bf x}-{\bf x'}|^{3}), \]
and
\begin{equation} G \ m_{np}^{2} \ = \ n_{{\varepsilon}_{0}} \ Q'_{M'} \ \Psi \ Q_{M} /(4 \ \pi) . \label{eq:gravitational_equivalence} \end{equation}
If small mass bodies are employed for both \( M' \) and \( M \) (as in the Cavendish apparatus), \( Q'_{M'} = Q_{M} = 1 \) and
\begin{equation} n_{{\varepsilon}_{0}} \ \Psi \ = \ 4 \ \pi \ G \  m_{np}^{2}. \label{eq:gravitational_equivalence_small} \end{equation}
This is a profound relationship that essentially redefines gravity in terms of collision parameters.
Instead of \( G \) being considered as an empirical constant, as in the Newton and Cavendish work, we have shown that it can be derived precisely from very fundamental properties of collisions and can actually be a function of \( n_{{\varepsilon}_{0}} \) that must depend on the stellar environment.
Using the numerical value of \( G \) obtained from the Cavendish apparatus:
\[ G \ = \ 6.67 \times 10^{-8} \ cm^{3} / (gram \ sec^{2}), \]
and employing
\[ m_{np} \ = \ 1.67 \times 10^{-24} \ gram, \]
\begin{equation} n_{{\varepsilon}_{0}} \ \Psi \ = \ 2.34 \times 10^{-54} \ external \ particles \ gram \ cm^{3} / sec^{2} . \label{eq:gravitational_equivalence_small_value} \end{equation}

Since the Newtonian field view and the external particle view predict different celestial orbits because of the apparent mass particle deficit associated with the external particle view (shown by example in equation~\ref{eq:number_deficit_example}), future improvement in level of precision in the independent mass measurements of the two bodies might eventually eliminate one of these explanations of gravity.
In the meantime, both views can be candidate theories as long as the error in the independent mass measurements is sufficiently large.
If the observed orbits of the two bodies are consistent with the Newtonian field view using present measured mass values and their error ranges, then the external particle view must have an apparent mass particle deficit that is less in magnitude than the estimated mass error, and thus an upper limit can be set on the size of the collision cross section.
If the mass error is expressed as a fraction \( E \) of the total mass measurement of the body,
\[ E \ > \ 3 \ \sigma_{\varepsilon,np,cc} \ n_{m'} \ R'/4, \]
or
\[ \sigma_{\varepsilon,np,cc} \ < \ 4 \ E/(3 \ n_{m'} \ R'). \]
If the earth and earth-moon are used, and if the mass particle distribution of both can be assumed to be uniform throughout, and if \( E = 0.05 \), \( n_{m'} \simeq 3.3 \times 10^{24} \ mass \ particles / cm^{3} \) (from an average material density of \(5 \ gram/cm^{3} \)), \( R' = 6.365 \times 10^{3} \ km \), then the requirement for viability of the external particle view is:
\begin{equation} \sigma_{\varepsilon,np,cc} \ < \ 3 \times 10^{-35} \ cm^{2}. \label{eq:sigma_upper_limit} \end{equation}
This merely sets an upper bound for viability of the theory.

Thus, for the viability of this model, the external particle responsible for gravity must have a \emph{low} but \emph{finite} probability of collision with all those bodies for which we have accurate independent mass measurements and for which the observed orbits are known to conform to the mass product relationship of the gravitational field orbit mechanics.
We do not expect that any feasible increases in accuracy of mass content will actually validate one view over the other because the capture and scattering collision cross sections we adopt later in this paper (for other reasons) are far smaller than \( 10^{-35} \ cm^{2} \) and would result in far smaller error terms.

All derivations so far have treated the positions of all mass bodies as stationary.
If relative motion were to be considered between the two bodies and between each of the bodies and the rest frame of the external particles, the calculations would have to be modified to take into consideration the body positions at retarded times determined by the propagation speed of the gravitational message in the medium created by the moving external particles.
These corrections would involve a power series expansion of terms such as \( ({\bf x}-{\bf x'}) \), \( v_{{\varepsilon}_{0}} \), and the velocities of both bodies.
We do not intend to incorporate this effect in this treatise because of the complexities involved in sorting out other first and second-order effects and in preparing to focus on very high external particle speeds, where other important relativistic effects must also be considered.
However, these retardation corrections necessarily have to be considered in any serious orbital calculations and may lead to important predicted and measurable differences that help to validate one theory of gravity over another.

\subsection{Expected Problems Arising from the Classical \newline Treatment of Collisions Between the Undisturbed \newline External Particles and a Test Body}

Using a non-relativistic approach in this entire section, we calculate the net first-order external particle flux as experienced by a test body moving with respect to the rest frame of the external particles.
We also calculate the corresponding net second-order momentum transfer (in the rest frame of the moving body) from those external particles that collide with this small idealized mass body.
The idealized body is again used here in order to simplify the calculations while retaining important first or second-order effects.
Lastly, we calculate the excess energy that is obtainable through collisions from the omnidirectionally impinging external particles and that may or may not accumulate in the mass body.
Since this excess energy is a first-order effect, a mass body having zero velocity relative to the external particle rest frame can be employed.
These calculations are done in order to compute the magnitudes of drag forces and possible energy accumulation that must be considered in an external particle view but that do not have to be considered at all in either the Newtonian or graviton view.
(For relativistic external particles, these calculations are not valid, but they nevertheless serve to identify very important issues.)
Other adjustments appropriate to the complete details of the interactions that occur for a specific type of external particle (when designated) must also be considered.

\subsubsection{Net External Particle Flux Experienced by a Moving \newline Reference Frame}

If we assume an omni-directional external particle distribution function relative to the rest frame of the external particles and consider a different reference frame moving at a small constant velocity \( -v_{M} \hat{\bf z} \) relative to that external particle rest frame, the external particle distribution function can be expressed relative to the moving reference frame as:
\[ f'_{\varepsilon}(x,v) \ = \ F'_{{\varepsilon}_{0}} \delta [v_{r}-(v_{{\varepsilon}_{0}}+v_{M}\sin \theta)], \]
where (\( \theta = \pi/2 \)) lies parallel to the \( \hat{\bf z} \) direction.
This leads to
\[ n_{{\varepsilon}_{0}} \ = \ \int F'_{{\varepsilon}_{0}} \ \delta [v_{r}-(v_{{\varepsilon}_{0}}+v_{M}\sin \theta)] \ \cos \theta \ d\theta \ d\phi \ v_{r}^{2} \ dv_{r} \ = \]
\[ 2 \ \pi \ F'_{{\varepsilon}_{0}} \ \int \ [v_{{\varepsilon}_{0}}^{2} \ + \ 2v_{{\varepsilon}_{0}}v_{M}\sin \theta \ + \ v_{M}^{2}\sin^{2} \theta] \ \cos \theta \ d\theta \ = \]
\begin{equation} 4 \ \pi \ F'_{{\varepsilon}_{0}} \ [v_{{\varepsilon}_{0}}^{2} \ + \ v_{M}^{2}/3]. \end{equation}
The corresponding net flux in the moving reference frame is:
\[ n_{\varepsilon}(x) \langle {\bf v}_{\varepsilon}(x) \rangle \ = \ \int F'_{{\varepsilon}_{0}} \ \delta [v_{r}-(v_{{\varepsilon}_{0}}+v_{M}\sin \theta)] \ v_{r} \ \hat{\bf v}_{r} \ \cos \theta \ d\theta \ d\phi \ v_{r}^{2} \ dv_{r}. \]
From symmetry, integration over \( v_{r} \) and \( \phi \) leaves only a surviving \( \hat{\bf z} \) component:
\[ 2 \ \pi \ F'_{{\varepsilon}_{0}} \ \hat{\bf z} \ \int \ [v_{{\varepsilon}_{0}}^{3} \ + \ 3v_{{\varepsilon}_{0}}^{2}v_{M}\sin \theta \ + \ 3v_{{\varepsilon}_{0}}v_{M}^{2}\sin^{2} \theta \ + \ v_{M}^{3}\sin^{3} \theta] \ \sin \theta \ \cos \theta \ d\theta \ = \]
\[ 4 \ \pi \ F'_{{\varepsilon}_{0}} \ \hat{\bf z} \ [v_{{\varepsilon}_{0}}^{2}v_{M} \ + \ v_{M}^{3}/5]. \]
Thus,
\begin{equation} n_{\varepsilon}(x) \langle {\bf v}_{\varepsilon}(x) \rangle \ = \ - \ n_{{\varepsilon}_{0}} \ {\bf v}_{M} \ [v_{{\varepsilon}_{0}}^{2} \ + \ v_{M}^{2}/5] \ / \ [v_{{\varepsilon}_{0}}^{2} \ + \ v_{M}^{2}/3]. \end{equation}

\subsubsection{Treatment of Elastic Collisions and Average Loss of Incident Particle Speed and Kinetic Energy}

If we consider an elastic collision between a nucleon that is initially at rest and a moving external particle that scores a direct hit, the velocity imparted to the nucleon is
\[ {\bf v}_{np,dh} \ = \ {\bf v}_{{\varepsilon}_{0}} \ [ 2 m_{\varepsilon} / (m_{np} + m_{\varepsilon}) ], \]
the post-collision velocity of the external particle is
\[ {\bf v}_{{\varepsilon}_{1,dh}} \ = \ - \ {\bf v}_{{\varepsilon}_{0}} \ [ (m_{np} -  m_{\varepsilon}) / (m_{np} + m_{\varepsilon}) ], \]
and the kinetic energy transferred to the nucleon is
\[ (1/2) m_{np} v_{np,dh}^{2} \ = \ (1/2) m_{\varepsilon} v_{{\varepsilon}_{0}}^{2} \ [4 m_{np} m_{\varepsilon} / (m_{np} + m_{\varepsilon})^2 ]. \]
We identify the (nucleon/external particle) energy transfer ratio for the direct hit as
\[ \beta_{dh} \ \equiv \ m_{np} v_{np,dh}^{2} / (m_{\varepsilon} v_{{\varepsilon}_{0}}^{2}) \ = \ 4 m_{np} m_{\varepsilon} / (m_{np} + m_{\varepsilon})^2. \]
The ``average'' scattering collision transfers a smaller amount of energy to the nucleon, the ratio of which is denoted by
\begin{equation} \beta_{av} \ \equiv \ \langle m_{np} v_{np,av}^{2} \rangle / [ m_{\varepsilon} v_{{\varepsilon}_{0}}^{2} ]. \end{equation}
\[ \langle m_{np} v_{np,av}^{2} \rangle \ + \ \langle m_{\varepsilon} v_{{\varepsilon}_{1,av}}^{2} \rangle \ = \ m_{\varepsilon} v_{{\varepsilon}_{0}}^{2}. \]
\[ m_{\varepsilon} v_{{\varepsilon}_{0}}^{2} \ \beta_{av} \ + \ \langle m_{\varepsilon} v_{{\varepsilon}_{1,av}}^{2} \rangle \ = \ m_{\varepsilon} v_{{\varepsilon}_{0}}^{2}. \]
\[ \langle v_{{\varepsilon}_{1,av}}^{2} \rangle / v_{{\varepsilon}_{0}}^{2} \ = \ (1 \ - \ \beta_{av}) \ > \ (1 \ - \ \beta_{dh}) \ = \ ( m_{np} - m_{\varepsilon} )^{2}/ ( m_{np} + m_{\varepsilon} )^{2} . \]
\begin{equation} \langle v_{{\varepsilon}_{1,av}} \rangle / v_{{\varepsilon}_{0}} \ > \ ( m_{np} - m_{\varepsilon} ) / ( m_{np} + m_{\varepsilon} ) . \end{equation}
Therefore, if \( m_{\varepsilon} \ll m_{np} \), \( v_{{\varepsilon}_{1,av}} \simeq v_{{\varepsilon}_{0}} \).

\subsubsection{The Problematic Drag Force on a Moving Body}

Throughout the volume of a small test body \( M \) that is at rest in the moving reference frame considered above in section 2.5.1, the number density of mass particles and the external particle distribution can be assumed to be uniform with respect to \( {\bf x} \).
For capture and scattering collisions between the external particles and these mass particles, the corresponding time rate of collisions throughout the entire body is:
\[ \int F'_{{\varepsilon}_{0}} \delta [v_{r}-(v_{{\varepsilon}_{0}}+v_{M}\sin \theta)] v_{r} \cos \theta d\theta d\phi v_{r}^{2} dv_{r} (\sigma_{\varepsilon,np,cc} + \sigma_{\varepsilon,np,sc,tnf}) \int n_{m} d^{3}x . \]

For capture collisions, the momentum transferred from an external particle to a mass particle is \( m_{\varepsilon} {\bf v}_{r} \), and for elastic scattering, the ``average'' momentum transferred is \( \hat{\bf v}_{r} ( m_{\varepsilon} v_{r} - m_{\varepsilon} \langle v'_{r} \cos \vartheta \rangle ), \)
where \( \vartheta \) is the scattering angle measured relative to \( \hat{\bf v}_{r} \).

The total time rate of momentum transfer to \( M \) is approximately:
\[ \int F'_{{\varepsilon}_{0}} \delta [v_{r}-(v_{{\varepsilon}_{0}}+v_{M}\sin \theta)] v_{r} m_{\varepsilon} v_{r} \hat{\bf v}_{r} \cos \theta d\theta d\phi v_{r}^{2} dv_{r} (\sigma_{\varepsilon,np,cc} + \sigma_{\varepsilon,np,sc,tpf}) N_{M} . \]
Incorporating symmetry about \( \hat{\bf z} \), this becomes: \pagebreak
\[ 2 \ \pi \ F'_{{\varepsilon}_{0}} \ m_{\varepsilon} \ \hat{\bf z} \ \sigma_{\varepsilon,np,cc} \ N_{M} \ \int \ \cos \theta \ d\theta \ ( v_{{\varepsilon}_{0}}+v_{M}\sin \theta )^{4} \ \sin \theta \ + \]
\[ 2 \ \pi \ F'_{{\varepsilon}_{0}} \ m_{\varepsilon} \ \hat{\bf z} \ \sigma_{\varepsilon,np,sc,tpf} \ N_{M} \ \int \ \cos \theta \ d\theta \ ( v_{{\varepsilon}_{0}}+v_{M}\sin \theta )^{4} \ \sin \theta \]
\[ = \ 2 \ \pi \ F'_{{\varepsilon}_{0}} \ m_{\varepsilon} \ \hat{\bf z} \ ( \sigma_{\varepsilon,np,cc} \ + \ \sigma_{\varepsilon,np,sc,tpf} ) \ N_{M} \ \cdot \]
\[ \int \ \sin \theta \ \cos \theta \ d\theta \ [ v_{{\varepsilon}_{0}}^{4} \ + \ 4 v_{{\varepsilon}_{0}}^{3} v_{M} \sin \theta \ + \ 6 \ v_{{\varepsilon}_{0}}^{2} v_{M}^{2} \sin^{2} \theta \ + \ 4 \ v_{{\varepsilon}_{0}} v_{M}^{3} \sin^{3} \theta \ + \ v_{M}^{4} \sin^{4} \theta ] \] 
\[ = - (4/3) n_{{\varepsilon}_{0}} \ [1 + 3 v^{2}_{M}/(5 v_{{\varepsilon}_{0}}^{2})]/[1 + v^{2}_{M}/(3 v_{{\varepsilon}_{0}}^{2})] \ m_{\varepsilon} v_{{\varepsilon}_{0}} {\bf v}_{M} ( \sigma_{\varepsilon,np,cc} + \sigma_{\varepsilon,np,sc,tpf} ) N_{M} \] 
\begin{equation} \simeq \ - \ (4/3) \ n_{{\varepsilon}_{0}} \ m_{\varepsilon} \ v_{{\varepsilon}_{0}} \ {\bf v}_{M} \ ( \sigma_{\varepsilon,np,cc} \ + \ \sigma_{\varepsilon,np,sc,tpf} ) \ N_{M} . \label{eq:classical_drag_force} \end{equation}

We evaluate this drag effect on the earth as it orbits the sun and assume that the sun is in the rest frame of the external particles (making the earth's speed relative to that rest frame constant).
We calculate the ratio of the earth's orbital momentum to the drag-induced time rate of change of that orbital momentum in order to produce a time estimate of how long it takes to substantially destroy this near-circular orbit.
From above,
\[ M \ dv_{M}/dt \ \simeq \ - \ (4/3) \ n_{{\varepsilon}_{0}} \ m_{{\varepsilon}_{0}} \ v_{{\varepsilon}_{0}} \ v_{M} \ ( \sigma_{\varepsilon,np,cc} \ + \ \sigma_{\varepsilon,np,sc,tpf} ) \ N_{M} \ . \]
Since this has the form of an exponentially decreasing speed, we calculate a time constant characteristic of the practical lifetime of the orbit:
\begin{equation} \tau_{drag} \equiv M v_{M} / [M (dv_{M}/dt)] \sim 3 m_{np} / [ 4 n_{{\varepsilon}_{0}} m_{\varepsilon} v_{{\varepsilon}_{0}} ( \sigma_{\varepsilon,np,cc} + \sigma_{\varepsilon,np,sc,tpf} )]. \label{eq:tau_drag} \end{equation}
If the external particle can be assumed to have much less mass than a nucleon, \( \Psi \) (from equation~\ref{eq:momentum_flux_coupling_term}) becomes
\begin{equation} \Psi \ \simeq \ v_{{\varepsilon}_{0}} \sigma_{\varepsilon,np,cc} m_{{\varepsilon}_{0}} v_{{\varepsilon}_{0}} ( \sigma_{\varepsilon,np,cc} + \sigma_{\varepsilon,np,sc,tpf} ), \label{eq:momentum_flux_coupling_term_small} \end{equation}
and, from equation~\ref{eq:gravitational_equivalence_small_value},
\[ n_{{\varepsilon}_{0}} m_{{\varepsilon}_{0}} v_{{\varepsilon}_{0}} ( \sigma_{\varepsilon,np,cc} + \sigma_{\varepsilon,np,sc,tpf} ) \simeq \]
\[ 2.34 \times 10^{-54}/( v_{{\varepsilon}_{0}} \sigma_{\varepsilon,np,cc} ) \ external \ particles \ gram / sec. \]
So,
\begin{equation} \tau_{drag} \ \simeq \ 1.5 \times 10^{40} \ \sigma_{\varepsilon,np,cc} \ ( v_{{\varepsilon}_{0}} / c ) \ sec. \end{equation}
Employing the conservatively large value of \( \sigma_{\varepsilon,np,cc} \) from equation~\ref{eq:sigma_upper_limit} and using \( v_{{\varepsilon}_{0}} = c \), this is less than 2 days, which would never allow a stable orbit!

An external particle, to be viable, must possess some special property to allow it to circumvent this large predicted drag force.
This will be treated in section 3.3.
  
\subsubsection{The Problematic Excess Collision Energy that may \newline Accumulate in a Body}

The time rate of increase of excess energy obtained from external particles colliding with a body \( M \) can be calculated similarly to equation~\ref{eq:collision_rate_per_vol_per_vol} by instead employing \( ( \sigma_{\varepsilon,np,cc} + \sigma_{\varepsilon,np,sc,tnf} ) \) and multiplying by the average energy made available from the colliding external particles.
Ignoring any relative body velocity and integrating over the volume of \( M \) and over external particle velocity space, the total time rate of increase of available energy is:
\[ dE_{excess}/dt \ = \]
\[ \int f_{\varepsilon} (x', v_{r}) v_{r} \cos \theta d\theta d\phi v_{r}^{2} dv_{r} \ (1/2) \ m_{{\varepsilon}_{0}} v_{r}^{2} ( \sigma_{\varepsilon,np,cc} + \beta_{av} \sigma_{\varepsilon,np,sc,tnf} ) \int n_{m} d^{3}x \] 
\[ = \ n_{{\varepsilon}_{0}} \ v_{{\varepsilon}_{0}} \ (1/2) \ m_{{\varepsilon}_{0}} \ v_{{\varepsilon}_{0}}^{2} \ (  \sigma_{\varepsilon,np,cc} + \beta_{av} \sigma_{\varepsilon,np,sc,tnf} ) \ N_{M} \]
\[ \simeq \ n_{{\varepsilon}_{0}} \ v_{{\varepsilon}_{0}} \ (1/2) \ m_{{\varepsilon}_{0}} \ v_{{\varepsilon}_{0}}^{2} \  \sigma_{\varepsilon,np,cc} \ N_{M} \]
because \( \beta_{av} \ll 1.0 \).
Assuming that this excess energy is captured by the earth from the external particles impinging from all directions, and ignoring any effects of motion relative to the rest frame of the external particles, we calculate the ratio of the earth's orbital kinetic energy (\( KE_{earth} \)) to the time rate of increase of this excess energy in order to produce a time estimate of how long it takes to generate the equivalent of the earth's orbital kinetic energy from collisions with external particles.
\begin{equation} \tau_{excess} \ \equiv \ KE_{earth} / (dE_{excess}/dt) \ \simeq \ m_{np} \ v_{M}^{2} / [ n_{{\varepsilon}_{0}} \ v_{{\varepsilon}_{0}} \ m_{{\varepsilon}_{0}} \ v_{{\varepsilon}_{0}}^{2} \ \sigma_{\varepsilon,np,cc} ]. \end{equation}
From equation~\ref{eq:momentum_flux_coupling_term_small}, 
\[ [ n_{{\varepsilon}_{0}} \ v_{{\varepsilon}_{0}} \ m_{{\varepsilon}_{0}} \ v_{{\varepsilon}_{0}}^{2} \ \sigma_{\varepsilon,np,cc} ] \ \simeq \ n_{{\varepsilon}_{0}} \ \Psi \ v_{{\varepsilon}_{0}} / ( \sigma_{\varepsilon,np,cc} + \sigma_{\varepsilon,np,sc,tpf} ) \]
Substituting \( n_{{\varepsilon}_{0}} \Psi \) from equation~\ref{eq:gravitational_equivalence_small_value},
\begin{equation} \tau_{excess} \ \simeq \ 2 \times 10^{32} \ ( \sigma_{\varepsilon,np,cc} + \sigma_{\varepsilon,np,sc,tpf} ) \ ( v_{{\varepsilon}_{0}} / c )^{-1} \ sec. \label{eq:tau_excess} \end{equation}
Even for relatively slow external particles, this is less than 1 second!

Relativistic effects, the extremely short interaction time intervals for relativistic external particles, and the details of the collisions for the particular type of external particle, however, could influence these results dramatically.
(This will be further discussed in section 3.2.)
Whether or not this excess energy is actually transferred to the nucleon and, furthermore, whether it accumulates in some form depend on the details of the collision and any collision-induced after-effects.

\section{Evaluation of the Neutrino as the Gravity-Bearing External Particle Based on Recent Neutrino Mass Data}

Owing to the recent discoveries regarding the neutrino and its established finite mass, the neutrino should be evaluated as a favorable candidate external particle to carry the gravity phenomenon.
The neutrino was in fact considered to be related to gravity by the previous work of Dirac and Gamow and noted to have in interaction the favorable spin of 2 [5].
Scientific literature data, developed in the 1950's, indicates for neutrinos an interaction scattering cross section with neutrons and protons of approximately \( 10^{-43} \ cm^{2} \) [6].
This scattering cross section was theoretically developed at a time when the neutrino was believed to be massless energy and in that sense treated as a field rather than as a particle.
Because the neutrino is now believed to have a finite mass, the value of its scattering cross section should be re-examined.

\subsection{Recent Neutrino Data}

The reported data for the neutrino are [7]: \\
\( v_{{\varepsilon}_{0}}/c \ \sim \ 0.95 \) ; \\
upper limits for the neutrino masses: \\
\hspace*{.5in} electron neutrino \( m < 5 \ eV \) ; \\
\hspace*{.5in} muon neutrino \( m < 170 \ KeV \) ; \\
\hspace*{.5in} tau neutrino \( 1 \ MeV < m < 18.2 \ to \ 30 \ MeV \) ; \\
Current information indicates:\\
\hspace*{.5in} electron neutrinos are non-showering; \\
\hspace*{.5in} muon neutrinos are showering; \\
\hspace*{.5in} tau neutrinos are at least partially showering. \\
Some interpretations of data suggest that all three types of neutrinos are present in a mixed state.

Experiments in Japan, using the Super-Kamiokande tank, detected only neutrino effects corresponding to a mass of \( 0.1 \) to \( 1 \ eV \).
However, this experiment was not designed to detect the tau neutrino [8].
The result of the Japanese experiment showing the scintillation of one water molecule out of fifty thousand tons of water by a single solar neutrino does not detract from the above postulates because the theory depends on an omni-directional external flux which is \emph{not} characterized solely by electron and muon neutrinos from the sun.
  
Our theory requires a substantial contribution from showering neutrinos.
Factual validation of the theory also requires that our sun must be a net sink rather than a source for these momenta-carrying (and transferring) external particles.
Recent data with respect to the decay of solar neutrinos indicate a decay time of less than 8 minutes;
additionally, a substantial proportion of higher-energy muon and tau neutrinos originate in super-nova explosions and larger and hotter stars rather than in our own sun.
Both of these observations support the above requirement.

The requirements from equation~\ref{eq:gravitational_equivalence_small_value} as pertains to the muon \( ( \mu ) \) and tau \( ( \tau ) \) neutrinos are tabulated in the following:
\[ n_{\mu} \ \sigma_{\mu,np,cc} \ ( \sigma_{\mu,np,cc} + \sigma_{\mu,np,sc,tpf} ) \ (100 \ KeV) \ (v_{\mu}/c)^{2} \ \simeq \]
\[ 2.34 \times 10^{-54} \ neutrinos \ gram \ cm^{3} / sec^{2} \]
\[ n_{\mu} \ \sigma_{\mu,np,cc} \ ( \sigma_{\mu,np,cc} + \sigma_{\mu,np,sc,tpf} ) \ \simeq \ (2.34/1.6) \times 10^{-47} \ neutrinos \ cm \]
\[ n_{\tau} \ \sigma_{\tau,np,cc} \ ( \sigma_{\tau,np,cc} + \sigma_{\tau,np,sc,tpf} ) \ (20 \ MeV) \ (v_{\tau}/c)^{2} \ \simeq \]
\[ 2.34 \times 10^{-54} \ neutrinos \ gram \ cm^{3} / sec^{2} \]
\[ n_{\tau} \ \sigma_{\tau,np,cc} \ ( \sigma_{\tau,np,cc} + \sigma_{\tau,np,sc,tpf} ) \ \simeq \ 2.34/3.2 \times 10^{-49} \ neutrinos \ cm \]

We proceeded in the following manner to adopt a range of neutrino parameters for subsequent analyses.
In the product term in equation~\ref{eq:gravitational_equivalence_small} we started with a value (from ref 6) for \( \sigma_{\nu,np,cc} \) and for \( \sigma_{\nu,np,sc,tpf} \) of \( 10^{-43} \ cm^{2} \).
This yielded a high energy density of \( 2.3 \times 10^{32} \ erg / cm^{3} \).
The above data on the \( \tau \) neutrino suggested that high energy neutrinos could be available in sufficient quantity that if postulated as the gravity-bearing particle would reduce the energy density requirement (because \( \sigma_{\nu,np,sc,tpf} \) is an almost linear function of neutrino energy using a massless model).

We therefore elected to employ the \( 20 \ MeV \ \tau \) neutrino as the subject particle.
Recalculating \( \sigma_{\nu,np,sc,tpf} \) for such an energetic neutrino yields in an analogous elastic scattering computation (following Euwema [6]) for a massless neutrino of \( 9 \times 10^{-40} \ cm^{2} \).
We assume that a similar approach for capture will yield a smaller capture cross section, but that this capture cross section will have to be increased when precise data on the neutrino mass are considered.
This presents a considerable variance.
The logic expressed above led us to employ a value of \( 10^{-40} \ cm^{2} \) (as a middle value) for the nucleon-neutrino capture cross section.
(Neutrino-neutrino collision cross sections for neutrinos of such energy are estimated to be \( \sigma_{\nu, \nu} \ \sim \ 10^{-60} \ cm^{2} \).)
We therefore commence to use the neutrino symbol \( \nu \) instead of the more general \( \varepsilon \) when describing quantities associated with the external particles.
Because the neutrino is neutral, has a small mass relative to the nucleon (ratio \( < 0.01 \)), and its speed is essentially unchanged in scattering collision, its scattering would contribute very little (compared to capture collisions) toward the creation of a gravitational influence by a gravitational mass body.
Scattered neutrinos do, however, likely contribute to momentum transfer. 
We believe that the following table 1 illustrates (in cgs units) related data concerning the practical range of \( \sigma_{\nu,np,cc} \) values for the range (\( 10^{-41} \ cm^{2} \le \sigma_{\nu,np,cc} \le 10^{-39} \ cm^{2} \)).

\vspace{0.2in}
\begin{tabular}{|l|l|l|l|} \hline
\( \sigma_{\nu,np,cc} \) & \( 10^{-41} \) & \( 10^{-40} \) & \( 10^{-39} \) \\ \hline
\( n_{{\nu}_{0}} \) & \( 3.7 \times 10^{32} \) & \( 3.7 \times 10^{30} \)& \( 3.7 \times 10^{28} \) \\ \hline
\( \nu \) energy density & \( 1.2 \times 10^{28} \) & \( 1.2 \times 10^{26} \)& \( 1.2 \times 10^{24} \) \\ \hline
\( \nu \) mass density & \( 1.3 \times 10^{7} \) & \( 1.3 \times 10^{5} \)& \( 1.3 \times 10^{3} \) \\ \hline
\( \nu \)-np collision freq. & \( 1.1 \times 10^{2} \) & \( 1.1 \times 10^{1} \)& \( 1.1 \times 10^{0} \) \\ \hline
\( \nu \)-\( \nu \) collision freq. & \( 1.1 \times 10^{-17} \) & \( 1.1 \times 10^{-19} \)& \( 1.1 \times 10^{-21} \) \\ \hline
\end{tabular} 

\subsection{Neutrino Collisions with a Body and the Disposition of Excess Energy}

The classical calculation of \( \tau_{excess} \) in equation~\ref{eq:tau_excess} yields for most neutrinos the value of about \( 2 \times 10^{32} \ ( \sigma_{\nu,np,cc} + \sigma_{\nu,np,sc,tpf} ) \ sec \), implying an intolerably rapid generation of available energy in any body.
However, the full details of neutrino collisions with nucleons involve the engendering of other nuclear endothermically generated particles which immediately consume and transport away energy.
The neutrino capture is only the beginning of such an interaction, and any subsequent stages of the interaction further dissipate energy.
Key to this understanding is that there must be no net accumulation of neutrinos in any mass body.

The term \( \partial n_{\varepsilon}(x')/\partial t \) in the flux continuity equation (~\ref{eq:continuity}) is basically a time rate of collisions rather than a rate of accumulation of external particles that are captured by collisions with nucleons.
We therefore postulate that the secondary particles (neutrinos, antineutrinos, etc.) that carry off any portions of the excess energy must have a much smaller collision cross section than that of the incident neutrinos (otherwise, the gravitational processes in \( M' \) would be incapable of inducing any net momentum transfer to a nearby body).
This reduction in cross section may result from different spin characteristics of the secondary particles.
We, therefore, do not currently consider these newly-spawned particles in the neutrino flux term as relates to gravitational influence (their only useful role has been to prevent a catastrophic accumulation of energy in any gravitational mass body).

There is no conflict between the conservation of total vector momentum (and even of energy) in the neutrinos and the ability of the neutrino distribution to still exert a gravitational influence on other distant bodies.
If a captured neutrino results in no emitted particles, there is an obvious momentum flux deficit in the neutrino distribution that propagates outward from the first body.
The gravitational effect on other distant bodies is critically dependent on this momentum flux deficit because it yields an imbalance in the local momentum flux to which those distant bodies are exposed, and they are pushed toward the first body.
To the extent that secondary particles are emitted by the first body so that it can maintain a constant energy, those secondary particles may be thought to remove or at least partially compensate the momentum flux deficit.
However, we postulated above that secondary particles do not possess the same properties as the original captured neutrino and thus present a much smaller collision cross section for nucleons.
Thus, they may carry momentum and energy, but it is not a ``useful'' momentum since such particles are much less likely to interact with other nucleons.
This implies that the original deficit developed instantaneously at the time of collision in the propagating neutrino momentum flux essentially remains unaffected by secondary emissions as concerns gravitational effects on other bodies.
 
\subsection{Elimination of the Neutrino Drag Force}

The classically derived drag force (see equation~\ref{eq:classical_drag_force}) that acts on a body moving at a uniform velocity relative to a uniform external particle distribution must, according to our theory, submit to some argument that eliminates that drag force.
A viable argument is that the subject external particle has relativistic speeds and therefore motion relative to its rest frame is not detectable.
This means that with neutrino speeds of \( c \), and employing the relativistic method for addition of velocities, the nucleons see no essential differences in relative speed between neutrinos approaching from any direction [9].
This requires that all neutrinos capable of significant gravitational momentum transfer have speed \( c \).

The physical justification for this is that neutrinos are born with speed \( c \) and have virtually no possibility of interacting with one another.
The theoretical cross section for neutrino-neutrino collisions \( \sim 10^{-60} \ cm^{2} \) leads to neutrino travel being undisturbed by other neutrinos for times at least as long as the lifetime of the universe (see Table 1)! 

\section{Development of a Neutrino-Based Gravitational Theory}

\subsection{Recasting of Gravitational Expressions in Terms of Neutrino Parameters}

Since previous discussions concerning the neutrino have introduced several simplifications (most important of which are \( v_{{\nu}_{1}} = v_{{\nu}_{0}} = c \) and \( m_{{\nu}_{2}} = m_{{\nu}_{1}} = m_{{\nu}_{0}}, \)), we incorporate them in the following to re-state and simplify several key expressions and note any new significance. \newline

Neutrino number flux divergence:
\begin{equation} {\bf \nabla}' \ {\bf \cdot} \ [ \ n_{\nu}(x') \langle {\bf v}_{\nu}(x') \rangle \ ] \ = \ - \ n_{\nu}(x') \ c \ \sigma_{\nu,np,cc} \ n_{np}(x')  \label{eq:neutrino_flux_divergence} \end{equation}
This equation is a probabilistic statement of the divergence of the neutrino flux and asserts that incident gravitational neutrinos are removed from the external particle distribution because of nucleons that are present in a gravitational mass body.
Combining the flux continuity equation~\ref{eq:continuity} with the time rate of collisions per unit volume (equation~\ref{eq:collision_rate_per_vol_per_vol}) and with the properties of the neutrino has made it possible to express the divergence of the flux in terms of \( n_{\nu}(x') \) itself because the neutrino velocity distribution can be expressed as the product of a speed of essentially \( c \) and a variable angular intensity distribution.

Net neutrino number flux:
\begin{equation} n_{\nu}(x) \langle {\bf v}_{\nu}(x) \rangle \ = \ - \ n_{{\nu}_{0}} \ c \ \sigma_{\nu,np,cc} \ Q'_{M'_{f}} \ N_{M'_{f}} \ ({\bf x}-{\bf x'})/(4 \ \pi \ |{\bf x}-{\bf x'}|^{3}) \label{eq:neutrino_flux} \end{equation}

Neutrino momentum flux coupling term:
\begin{equation} \Psi_{\nu} \ = \ c \ \sigma_{\nu,np,cc} \ m_{{\nu}_{0}} \ c \ [ \sigma_{\nu,np,cc} \ + \ \sigma_{\nu,np,sc,tpf} ], \end{equation}

Time rate of change of momentum induced by the neutrino distribution on a small test body \( M_{s} \):
\[ d{\bf P}_{s}/dt \ = \]
\[ - [ n_{{\nu}_{0}} c \sigma_{\nu,np,cc} Q'_{M'_{f}} N_{M'_{f}} ({\bf x}-{\bf x'})/(4 \pi |{\bf x}-{\bf x'}|^{3}) ] \ [ m_{{\nu}_{0}} c (\sigma_{\nu,np,cc} + \sigma_{\nu,np,sc,tpf} ) N_{M_{s}} ] \]
\begin{equation} = \ [ \ n_{\nu}(x) \langle {\bf v}_{\nu}(x) \rangle \ ] \ [ \ m_{{\nu}_{0}} \ c \ (\sigma_{\nu,np,cc} + \sigma_{\nu,np,sc,tpf} ) \ N_{M_{s}} \ ]. \label{eq:neutrino_body_force} \end{equation}

The gravitational equivalence for the neutrino:
\begin{equation} n_{{\nu}_{0}} \ c \ \sigma_{\nu,np,cc} \ m_{{\nu}_{0}} \ c \ (\sigma_{\nu,np,cc} + \sigma_{\nu,np,sc,tpf} ) \ = \ 4 \ \pi \ G \  m_{np}^{2}. \label{eq:neutrino_gravitational_equivalence} \end{equation}

These facilitate the development of a straightforward gravitational potential.

\subsection{Neutrino-Based Gravitational Potential}

\subsubsection{Gravitational Properties of the Neutrino Flux}
 
In equation~\ref{eq:neutrino_flux_divergence} the divergence is proportional to the product of the nucleon density, local neutrino density, and capture collision cross section between the nucleon and the neutrino.
The net number flux term \( n_{\nu} (x) \langle v_{\nu} (x) \rangle \) in equation~\ref{eq:neutrino_flux} arises from a spatial change to the neutrino velocity distribution caused by the presence of the gravitational body \( M' \).

The quantity \( (\sigma_{\nu,np,cc} + \sigma_{\nu,np,sc,tpf} ) N_{M_{s}} \) in equation~\ref{eq:neutrino_body_force} for the aggregate of nucleons in the test body (rather than its total mass content) represents that body's contribution in permitting neutrino momentum to be transferred to \( M_{s} \).
The momentum of a test body is changed by gravity only if there is a \emph{net} neutrino flux.
We have previously asserted for the relativistic neutrinos that
\begin{equation} n_{\nu}(x') \langle {\bf v}_{\nu}(x') \rangle \ \equiv \ 0 \end{equation}
unless there exists nearby gravitational mass particles.
This assertion also involves probabilistic behavior, and it follows that only a nearby gravitational mass body can cause any net neutrino-induced gravitational effect on a test body by creating a nearby divergence in the neutrino distribution.
A gravitational mass body creates a non-zero component of this net flux only in the radial direction relative to itself (ignoring retardation effects due to high relative motion).
No net flux component (and no drag) is created perpendicular to its radial direction.
From the form of \( n_{\nu}(x) \langle {\bf v}_{\nu}(x) \rangle \) in equation~\ref{eq:neutrino_flux},
\begin{equation} {\bf \nabla} \ {\bf \times} \ [n_{\nu}(x) \langle {\bf v}_{\nu}(x) \rangle] \ \equiv \ 0, \end{equation}
and the net flux can be expressed in any region external to the gravitational mass body as the gradient of a scalar function:
\begin{equation} n_{\nu}(x) \langle {\bf v}_{\nu}(x) \rangle \ = \ n_{{\nu}_{0}} \ c \ \sigma_{\nu,np,cc} \ Q'_{M'} \ N_{M'} \ {\bf \nabla} \ [1/(4 \pi|{\bf x}-{\bf x'}|)]. \end{equation}

For a specific value of \( n_{{\nu}_{0}} \), the largest net neutrino flux that can possibly be generated results from equating the distribution function to zero in one half of the angular range to simulate what would result near the surface of a very large gravitational body that completely captures all neutrinos that penetrate its surface (especially over the non-local portion of the surface).
To calculate this, the approximation that was used in equation~\ref{eq:velocity_distribution_finite} is again employed, and the net flux resulting from integrating \( F_{{\nu}_{0}} \delta (v_{r}-c) {\bf v_{r}} \) over the angular range \( 0 \le \theta \le \pi/2 \) is canceled by the \( a'_{M'} \) term:
\[ \int_{0}^{\pi/2} \cos \theta \ d\theta \int_{0}^{2 \pi} d\phi \int v_{r}^{2} {\bf v_{r}} F_{{\nu}_{0}} \delta ( v_{r} - c ) \ dv_{r} \ - \]
\[ \int_{-\pi/2}^{\pi/2} \cos \theta \ d\theta \int_{0}^{2 \pi} d\phi \int  [ v_{r}^{2} {\bf v_{r}} a_{M'} \delta ( v_{r}, c, \theta,  \theta', \phi, \phi' )/(c^{2} r_{0}^{2}) ] \ dv_{r} \ = \ 0, \]
where \( r_{0} \) is the outer radius of the gravitational body.
Since only the \( \hat{\bf z} \) component of \( {\bf v}_{r} \) survives,
\[ \int_{0}^{\pi/2} \sin \theta \cos \theta \ d\theta \ 2 \pi \ c^{3} \ F_{{\nu}_{0}} \ = \ a_{M'} \ c \ / \ r_{0}^{2}. \]
Therefore,
\begin{equation} a_{M'} \ = \ n_{{\nu}_{0}} \ r_{0}^{2} \ / \ 4. \end{equation}
The maximum possible net flux is then
\begin{equation} [ \ n_{\nu}(x) \langle {\bf v}_{\nu}(x) \rangle \ ]_{max} \ = \ - n_{{\nu}_{0}} c \ {\bf {\hat r}}/4 \label{eq:maximum_neutrino_flux} \end{equation}
and is caused by the total absence of any outward-directed neutrinos at the surface of the body.
This maximum possible net flux is therefore limited by the value of \( n_{{\nu}_{0}} \) of the external neutrino distribution and by the ability of a large gravitational body to develop a dense nucleon region that effectively absorbs all neutrinos that enter it.

When the \( a_{M'} \delta ( v_{r}, c, \theta,  \theta', \phi, \phi' ) \) approximation term was employed in equation~\ref{eq:attenuated_velocity_distribution}, it was described as being inexact within the immediate vicinity of \( M' \).
This is evident because it does not eliminate the distribution magnitude over the entire angular range \( 0 \le (\theta - \theta') < \pi/2 \) as is actually accomplished by such a massive body.
However, quantities that rely on the integral of the distribution function over \( \theta \) rather than on the distribution function itself can be considered to be reliably calculated by using \( \delta ( v_{r}, c, \theta,  \theta', \phi, \phi' ) \) even in the near field region \( r_{0} < r < 2 r_{0} \).

\subsubsection{The Gravitational Potential and its Mapping}

The time rate of momentum change for test body \( M_{s} \) can now be written:
\[ d{\bf P}_{s}/dt \ = \]
\[ n_{{\nu}_{0}} c \sigma_{\nu,np,cc} Q'_{M'} N_{M'} m_{{\nu}_{0}} c (\sigma_{\nu,np,cc} + \sigma_{\nu,np,sc,tpf} ) N_{M_{s}} {\bf \nabla} \ [1/(4 \pi|{\bf x}-{\bf x'}|)] \]
\begin{equation} = \ - \ {\bf \nabla} \ \Phi_{s}({\bf x},{\bf x'}), \end{equation}
where
\[ \Phi_{s}({\bf x},{\bf x'}) \ \equiv \]
\begin{equation} - n_{{\nu}_{0}} c \sigma_{\nu,np,cc} Q'_{M'} N_{M'} m_{{\nu}_{0}} c ( \sigma_{\nu,np,cc} + \sigma_{\nu,np,sc,tpf} ) N_{M_{s}} /(4 \pi|{\bf x}-{\bf x'}|) \end{equation}
is a probabilistic potential.

It is useful to introduce the term \( \Phi_{np} \) to represent the potential per nucleon averaged over the aggregate (herein referred to as the averaged nucleon).
\[ \Phi_{np} \ \equiv \ \Phi_{s}/N_{M_{s}} \ = \]
\begin{equation} - \ n_{{\nu}_{0}} \ c \ \sigma_{\nu,np,cc} \ Q'_{M'} \ N_{M'} \  m_{{\nu}_{0}} \ c \ ( \sigma_{\nu,np,cc} + \sigma_{\nu,np,sc,tpf} ) /(4 \pi|{\bf x}-{\bf x'}|). \label{eq:potential_per_nucleon} \end{equation}
We define this as the fundamental gravitational potential for this probabilistic theory involving neutrino/nucleon collisions.
The equivalence of \( \Phi_{np} \) to a Newtonian gravitational potential for the mass of a single nucleon test body can be verified by substituting the value for \( n_{{\nu}_{0}} c \ \sigma_{\nu,np,cc} \ m_{{\nu}_{0}} c \ ( \sigma_{\nu,np,cc} + \sigma_{\nu,np,sc,tpf} ) \) from equation~\ref{eq:neutrino_gravitational_equivalence} and setting \( Q'_{M'} \ N_{M'} \) to \( 1 \).

\section{Recasting of Inertial Mass in Terms of the Neutrino Distribution}

The objectives of this section are to describe the behavior when a body moves or is accelerated relative to an inertial frame and to determine the meaning of the inertial mass with respect to both an accelerating body and the neutrino distribution.

A stationary observer in an inertial frame with no nearby gravitational bodies describes the neutrino distribution as being perfectly homogeneous and omnidirectional and characterized by \( F_{\nu} \delta (v_{r} - c) \).

Any local disturbance made to the neutrino distribution may be said to propagate outward and is here described to do so by virtue of the affected particles traveling independently to carry the momenta ``information''.

We now consider a stationary inertial observer and a small stationary gravitational test body that has been present for a long time (so that there is no time-dependent behavior).
This observer, like the first one, describes the neutrino background distribution as being perfectly homogeneous and omnidirectional and characterized by \( F_{\nu} \delta (v_{r} - c) \) except for whatever disturbance is superimposed by the body.
Such a body at rest has a continuous effect on the neutrinos by capturing a small portion of those that converge on and enter the body.
Each neutrino capture results in a loss of momentum and kinetic energy in a \emph{local} portion of the neutrino distribution.
If and when the capture event is later energy-compensated in the body by the emission of secondary particles, the total vector momentum of those particles is zero relative to the body, so the vector momentum impulse earlier captured by the body is never returned to the neutrino distribution by that neutrino or its secondary particles.
However, if other neutrino capture events are also taken into account, the net vector momentum change to the neutrinos (``neutrinos'' herein used collectively for all neutrinos) is zero owing to spherical symmetry.
Thus, averaged over small but finite increments of time, the body's vector momentum will not change and the net neutrino vector momentum will not change.
This is true regardless of secondary emission (but it could be possible for secondary emission to also maintain as constant the net energy of the body and the net kinetic energy of the neutrinos).
By reasoning similar to that used in equation~\ref{eq:collision_rate}, the total time rate of transfer of omnidirectional momentum from the neutrinos is
\begin{equation} n_{{\nu}_{0}} \ c \ m_{{\nu}_{0}} \ c \ (\sigma_{\nu,np,cc} + \sigma_{\nu,np,sc,tpf} ) \ N_{M}, \label{eq:omni-directional_momentum} \end{equation}
and
\[ n_{{\nu}_{0}} \ c \ m_{{\nu}_{0}} \ c \ \sigma_{\nu,np,cc} \ N_{M} \]
expresses a time rate of conversion of omnidirectional neutrino momentum by the body into a form of no useful gravitational consequence.

We now consider a \emph{different} stationary inertial observer watching a small gravitational body move by at a constant velocity \( {\bf v}_{M} \).
This observer, like the first one, describes the neutrino background distribution in his frame as being perfectly homogeneous and omnidirectional and characterized by \( F_{\nu} \delta (v_{r} - c) \) except for whatever disturbance is superimposed by the body.
The observer sees this moving body as acting similarly with regard to capturing neutrinos and later possibly emitting secondary particles, but there is a spatial and time dependent aspect to the propagation of the momentum flux deficit in the neutrino distribution that causes the observer to perceive that the net vector momentum of the neutrino distribution is increasing with time in the direction of the body velocity and in proportion to the magnitude of that body's velocity.
There is an observable moving pattern in the surrounding neutrino distribution that moves outward from the moving body and affects new regions of space as time progresses.
This pattern is a spherically expanding (but retarded) region of altered neutrino characteristics.
Low body velocity produces an almost concentric disturbance pattern, and high velocity produces what appears as a wave front. 
In the extreme, a body moving at speed \( c \) would create a disturbance pattern bounded by a moving cone whose apex moves with the body and whose leading edge is at a \( 45 \) degree angle with the axis of body motion.
Although this conical region will have a sharp transition in the momentum flux deficit, it is not likely to host a neutrino ``shock wave'' because of the extremely small neutrino-neutrino collision cross section.

Quantitatively, the stationary observer sees the body moving at (\( {\bf v}_{M} = - v_{M} \hat{\bf z} \)).
Neutrinos that approach the moving body traveling \emph{opposite} to the body's velocity appear to have a higher rate of impinging on the body because of the higher closure rate and appear to change momentum by a larger amount when captured because they must reverse direction and travel with the body.
Neutrinos that approach the moving body traveling in the \emph{same} direction as the body velocity appear to have a lower rate of impinging because of the lower rate of closure and appear to change momentum by a smaller amount when captured because they do not need to reverse direction.
As perceived by the stationary observer, the neutrino closure speed relative to the body is \( (c + v_{M} \sin \theta) \) where (\( \theta = \pi/2 \)) for neutrinos traveling in the \( \hat{\bf z} \) direction.
The momentum impulse transferred to the body from each colliding neutrino is \( (m_{{\nu}_{0}} c \hat{\bf z} \sin \theta + m_{{\nu}_{0}} v_{M} \hat{\bf z}) \).
This momentum impulse expression deliberately ignores all but the \( \hat{\bf z} \) component because the others sum to zero over finite time intervals (from symmetry of the neutrino distribution).
To the stationary observer, the average rate at which the neutrino distribution transfers momentum to the moving body is:
\[ (d{\bf P}_{M}/dt)_{(M \leftarrow \nu)} \ = \]
\[ \int d^{3}x \ \int \ F_{\nu} \ \delta (v_{r}-c) \ (c + v_{M} \sin \theta) \ (m_{{\nu}_{0}} c \ \hat{\bf z} \sin \theta + m_{{\nu}_{0}} v_{M} \hat{\bf z}) \ \cdot \]
\[ ( \sigma_{\nu,np,cc} + \sigma_{\nu,np,sc,tpf} ) \ n_{np}(x) \ \cos \theta \ d\theta \ d\phi \ v_{r}^{2} \ dv_{r} \ = \]
\[ 2 \ \pi \ F_{\nu} \ c^{2} \ m_{{\nu}_{0}} \ ( \sigma_{\nu,np,cc} + \sigma_{\nu,np,sc,tpf} ) \ N_{M} \ \hat{\bf z} \ \int \ \cos \theta \ d\theta \ (c \sin \theta + v_{M}) \ (c + v_{M} sin \theta) \]
\begin{equation} = \ - \ (4/3) \ n_{{\nu}_{0}} \ m_{{\nu}_{0}} \ c \ ( \sigma_{\nu,np,cc} + \sigma_{\nu,np,sc,tpf} ) \ N_{M} \ {\bf v}_{M}. \label{eq:perceived_force} \end{equation}
Since the body continues to move at a constant velocity as observed by the stationary observer, that observer tends to conclude that there must be some additional apparent force that keeps that body moving at constant velocity and, in turn, transfers net momentum to the neutrino distribution.
This apparent force that ultimately seems to cause the net rate of change of momentum of the neutrino distribution is \pagebreak
\[ {\bf Force}_{(\nu \leftarrow M)} \ = \ (d{\bf P}_{\nu}/dt)_{(\nu \leftarrow M)} \ = \]
\begin{equation} (4/3) \ n_{{\nu}_{0}} \ m_{{\nu}_{0}} \ c \ ( \sigma_{\nu,np,cc} + \sigma_{\nu,np,sc,tpf} ) \ N_{M} \ {\bf v}_{M}. \end{equation}
This net time rate of change of vector momentum, in magnitude, is much less than the time rate of omnidirectional consumption of momentum (from equation~\ref{eq:omni-directional_momentum}) by the ratio \( 4 | v_{M} | / (3 c) \).
The apparent force can be viewed as a product:
\begin{equation} (d{\bf P}_{\nu}/dt)_{(\nu \leftarrow M)} \ = \ ( dM_{\nu}/dt ) \ {\bf v}_{M}, \label{eq:mass_rate_of_capture} \end{equation}
where
\[ dM_{\nu}/dt = (4/3) \ n_{{\nu}_{0}} \ m_{{\nu}_{0}} \ c \ ( \sigma_{\nu,np,cc} + \sigma_{\nu,np,sc,tpf} ) \ N_{M} \]
is the time rate of capture of directed neutrino mass (from net vector momentum transfer rather than from omnidirectional momentum transfer) that then moves at \( {\bf v}_{M} \).
This velocity-related component of mass capture arises only because of the net neutrino vector momentum capture.
It is not observable in the rest frame of the body, so it can be considered to be an artifact of the body motion relative to the observer. 
No matter what the body velocity is, the neutrino velocity distribution as measured in the reference frame of the moving body appears to remain at (\( F_{{\nu}_{0}} \ \delta (v_{r}-c) \)).
The result is that, in regard to forces, we cannot rely on what different inertial observers report when they are \emph{not} at rest in the inertial frame in which the force is applied.

However, if we ask a stationary inertial observer to apply a constant force for a short time interval to a body initially at rest in his frame in order to impart to it an eventual constant velocity, we may rely on his observations if that imparted velocity is very small.
This stationary observer would report that he applied a real force (\( {\bf Force}_{(M \leftarrow app)} = - Force_{(M \leftarrow app)} \hat{\bf z} \)) during a small time interval \( \triangle t \) that on average resulted in a uniform body acceleration of (\( {\bf a}_{M} = - a_{M} \hat{\bf z} \)) during that time interval so that the final body velocity was (\( \triangle {\bf v}_{M} = - v_{M} \hat{\bf z} \)).
Furthermore, the observer would report that the average velocity of the body during that time interval was  \( \triangle {\bf v}_{M}/2 \) which resulted in a net neutrino momentum change of (\( (4/3) \ n_{{\nu}_{0}} \ m_{{\nu}_{0}} \ c \ ( \sigma_{\nu,np,cc} + \sigma_{\nu,np,sc,tpf} ) \ N_{M} \ (\triangle {\bf v}_{M}/2) \ \triangle t \)) during that time interval.
We conclude that
\begin{equation} {\bf Force}_{(M \leftarrow app)} \ = \ (2/3) \ n_{{\nu}_{0}} \ m_{{\nu}_{0}} \ c \ ( \sigma_{\nu,np,cc} + \sigma_{\nu,np,sc,tpf} ) \ N_{M} \ \triangle {\bf v}_{M} . \end{equation}

From the stationary observer's point of view, there are two reasons why the sum of such neutrino momentum impulses opposes the acceleration of the body:
the momentum transfer is larger for neutrinos traveling opposite to the body's velocity than for neutrinos traveling in the same direction.
The other is that the rate of distance-closure is higher for neutrinos traveling opposite to the body's velocity.
We have already calculated the appropriate integral expressing this as a continuum, but it is useful to write it as a sum of such impulses over the time interval \( \triangle t \): \pagebreak
\[ \triangle {\bf P}_{(M \leftarrow \nu)} \ = \ \sum_{i} {\bf p}_{(M \leftarrow \nu)_{i}} \ = \]
\begin{equation} (2/3) \ \hat{\bf z} \ n_{{\nu}_{0}} \ m_{{\nu}_{0}} \ c \ ( \sigma_{\nu,np,cc} + \sigma_{\nu,np,sc,tpf} ) \ N_{M} \ {\triangle v_{M}} \ \triangle t. \label{eq:sum_of_momentum_impulses} \end{equation}
Note that this is directed \emph{opposite} to the acceleration and that if \( {\triangle v_{M}} \) were \( 0 \), there would be no net momentum transfer during \( \triangle t \).

The general expression of Newton's second law is
\[ {\bf Force}_{(M \leftarrow app)} \ = \ (d{\bf P}_{M}/dt)_{M \leftarrow app}. \]
Concerning only a body whose number of nucleons is constant during the time interval, this becomes
\[ {\bf Force}_{(M \leftarrow app)} \ = \ M_{I} \ d{\bf v}_{M}/dt, \]
where \( M_{I} \) is the inertial mass of the body as utilized by Newton.
If we consider that the proper understanding of the above is that the external force tending to first accelerate the body is opposed by a set of responding neutrino momenta impulses tending to limit the acceleration, the balancing acceleration can be determined by summing:
\begin{equation} {\bf Force}_{(M \leftarrow app)} \ + \ {\bf Force}_{(M \leftarrow \nu)} \ = 0. \end{equation}
The force exerted by the neutrinos on the accelerating body is
\[ {\bf Force}_{(M \leftarrow \nu)} \ = \ \triangle {\bf P}_{(M \leftarrow \nu)}/\triangle t \ = \]
\begin{equation} (2/3) \ \hat{\bf z} \ n_{{\nu}_{0}} \ m_{{\nu}_{0}} \ c \ ( \sigma_{\nu,np,cc} + \sigma_{\nu,np,sc,tpf} ) \ N_{M} \ ({\triangle v_{M}}/\triangle t) \ \triangle t \end{equation}
and is generated only as a response to body acceleration.
Thus,
\begin{equation} M_{I} \ {\bf a}_{M} \ - \ {\bf a}_{M} \ (2/3) \ n_{{\nu}_{0}} \ m_{{\nu}_{0}} \ c \ ( \sigma_{\nu,np,cc} + \sigma_{\nu,np,sc,tpf} ) \ N_{M} \ \triangle t \ = \ 0 \end{equation}

The applied force actually causes the body to move in the direction of that force but in some undetermined manner.
As soon as the moving body starts to collide with neutrinos, however, those neutrinos resist the changing body motion in a statistically organized manner and establish an acceleration value through which the net neutrino force balances the applied force.
If the applied force had been larger, the greater body motion would have caused a higher neutrino resistive force by way of a higher momentum transfer rate, and a higher acceleration value would have resulted.
The resistive force is thus a stabilizing mechanism.
However, if the neutrino distribution were sparse, the resulting acceleration would have a large variance due to the time-separated neutrino momentum impulses.
 
A dimensional analysis of the above equation confirms that the \( \triangle t \) on the right side is necessary to provide the proper dimensions, but the equation cannot be left in this form because the magnitude of \( \triangle t \) is unspecified.
A resolution of this dilemma is obtained by noting that equation~\ref{eq:mass_rate_of_capture} has the same  expression \( [ n_{{\nu}_{0}} m_{{\nu}_{0}} c ( \sigma_{\nu,np,cc} + \sigma_{\nu,np,sc,tpf} ) N_{M} ] \) except for the numeric coefficient, and that term is interpreted as a time rate of neutrino mass capture (from net vector momentum capture) by the body.
Newton's inertial mass is therefore the magnitude of the vector summation of momentum impulses delivered to the body during some time interval and then divided by the product of the time interval (in seconds) and the average body speed during that time interval.
As such (and like thermodynamic pressure) it is an operational definition involving a summation over a time interval.
 
This inertial mass convention (and units) requires the above \( \triangle t \) to be equal to one second in order to determine the proper numerical value for the inertial mass:
\begin{equation} M_{I} \ \equiv \ (2/3) \ n_{{\nu}_{0}} \ m_{{\nu}_{0}} \ c \ ( \sigma_{\nu,np,cc} + \sigma_{\nu,np,sc,tpf} ) \ N_{M} \ neutrino \ gram/sec. \label{eq:inertial_mass} \end{equation}

To the inertial observer, this inertial mass is ficticious except while the body is being accelerated because it was not manifest at all when the body was first at rest, and would not be manifest to any inertial observer traveling with the body after it reaches its final velocity.
Furthermore, the inertial mass belongs to \emph{both} the body and to the neutrinos (rather than to either individually) because it incorporates \( \sigma_{\nu,np,cc} \) and \( \sigma_{\nu,np,sc,tpf} \), which are collision cross sections of a nucleon-neutrino pair.
We treat \( \triangle {\bf P}_{(M \leftarrow \nu)} \) (and \( M_{I} \)) as being physically non-zero only in the interval during which a non-zero external force is applied.
It should be noted from the way equation~\ref{eq:inertial_mass} defining inertial mass has been derived in this section (based on a postulate concerning the inability to detect inertial motion relative to the net rest frame of the neutrinos) that any mass quantity does not \emph {itself} have a relativistic correction.
Historically, length and time readings in one reference frame were correctly compared with length and time readings in a moving reference frame (even one with high relative velocity) to obtain relativistic corrections to those quantities.
However, the comparisons of inertial mass values in the different frames were not valid because no inertial observer was positioned in the rest frame of the \emph{mass} at any time.
Force measurements and their effect on inertial mass cannot be directly compared when high relative velocities are involved because net force itself implies a violation of inertial conditions.
We cited earlier how ``ficticious'' forces could be reported when observing a body with a finite relative velocity. 
Since the measuring apparatus cannot be even conceptually transported between the moving reference frames and since a single observer cannot simultaneously reside in two reference frames (one inertial and the other non-inertial), then successful comparisons involving inertial mass and force are not possible.
These considerations were what prompted us to analyze force and acceleration only for the case of infinitesimally small body velocities relative to the observer's inertial frame.
The \( \sigma_{\nu,np,cc} \) and \( \sigma_{\nu,np,sc,tpf} \) terms are areas \emph{transverse} to the direction of motion and would \emph{not} change relativistically.
The neutrinos have speed \( c \) and finite energy, and we have postulated that these values are the same for all inertial observers.
Therefore, no quantities in the derivation of \( M_{I} \) require relativistic corrections.
Einstein, in describing gravity waves, concluded that they must travel exactly at \( c \).

The above relationship (equation~\ref{eq:inertial_mass}) as applies to the inertial mass of a nucleon can be combined with the neutrino gravitational equivalence (equation~\ref{eq:neutrino_gravitational_equivalence}) to solve for the numerical value of \( \sigma_{\nu,np,cc} \): 
\[ n_{{\nu}_{0}} \ c \ \sigma_{\nu,np,cc} \ m_{{\nu}_{0}} \ c \ ( \sigma_{\nu,np,cc} + \sigma_{\nu,np,sc,tpf} ) \ = \]
\[ 4 \ \pi \ G \ m_{np} \ (2/3) \ n_{{\nu}_{0}} \ m_{{\nu}_{0}} \ c \ ( \sigma_{\nu,np,cc} + \sigma_{\nu,np,sc,tpf} ). \]
So,
\begin{equation} \sigma_{\nu,np,cc} \ = \ (8/3) \ \pi \ G \ m_{np} / c \ = \ 3.12 \times 10^{-41} \ cm^{2} \label{eq:derived_sigma} \end{equation}
This is an independently derived value that is within the range of values obtained from other theoretical calculations of the neutrino-nucleon cross section, and we will use this numerical value henceforth for the value of \( \sigma_{\nu,np,cc} \).
This leaves
\[ n_{{\nu}_{0}} \ m_{{\nu}_{0}} \ c^{2} \ ( \sigma_{\nu,np,cc} + \sigma_{\nu,np,sc,tpf} ) \ = \]
\begin{equation} 4 \ \pi \ G \ m^{2}_{np} / \sigma_{\nu,np,cc} \ = \ 7.55 \times 10^{-14} \ gram \ cm / sec^{2}. \end{equation}
If we assume that \( \sigma_{\nu,np,sc,tpf} \sim \sigma_{\nu,np,cc} \), the neutrino energy density can now also be estimated numerically from the above: 
\[ n_{{\nu}_{0}} \ m_{{\nu}_{0}} \ c^{2} \ \sim \ 2 \ \pi \ G \ m_{np}^{2}/ \sigma^{2}_{\nu,np,cc} \ = \]
\begin{equation} 1.2 \times 10^{27} \ erg / cm^{3}. \label{eq:neutrino_energy_density} \end{equation}
Henceforth, any numeric quantities developed using this approximation between \( \sigma_{\nu,np,sc,tpf} \) and \( \sigma_{\nu,np,cc} \) will be indicated by the use of \( \sim \).

\section{The Fundamentals that Give Rise to Newton's Laws of Motion and to Kinetic Energy \newline Conservation Under Special Conditions}

In the context of our treatment, when related to Newton's lws, we postulate that any initial motion induced by the impressed force is almost immediately opposed by a neutrino collisional force that increases in proportion to the magnitude of the acceleration and is directed opposite to it.
The resultant net force on the body is zero, and the net acceleration is determined by:
\[ {\bf Force}_{M \leftarrow app} \ = \ M_{I} \ d{\bf v}_{M}/dt \ = \]
\[ [(2/3) \ n_{{\nu}_{0}} \ m_{{\nu}_{0}} \ c \ ( \sigma_{\nu,np,cc} + \sigma_{\nu,np,sc,tpf} ) \ N_{M}] \ d{\bf v}_{M}/dt. \]
This net acceleration is a response to the superposition of two forces.
The neutrino response to acceleration is very stabilizing because it has almost unlimited capacity to transiently oppose any applied force, no matter how large. 

A neutrino-based theory contributes a better understanding of Newton's third law by asserting that the effect of an external applied force on a body almost immediately results in a change to the vector momentum of the surrounding neutrino distribution.
As such, it can really be considered to be an extension of the second law rather than an independent law.
We treat it as such but expand upon it in order to illuminate body-body collision phenomena when considered in the presence of the neutrino distribution.
As far as the body is concerned, there are two real forces being applied to it - the external force causing the initial motion and the [slightly lagging (discussed below)] force imposed by the neutrinos in resisting such body acceleration.
Rather than themselves being continuously accelerated, the neutrinos are momentarily captured by the body and impart a set of vector momentum impulses (with non-zero sum) back on the body.
This is a key distinction because it precludes relying on a ficticious body inertial mass force that resists and limits the acceleration (even for the neutrinos).

The result of this line of reasoning, when applied to two colliding bodies, is essentially a momentum conservation principle for the combined momenta of the two bodies.
The external force imposed on body 1 results from the deformation forces set up in body 2 (due to impact) and their transfer through the area of contact to body 1.
Body 1, which itself deforms, takes on a net acceleration which results from the mediation of the neutrinos.
Body 1 finds its own nucleons only able to accelerate at a rate determined by the matching of the deformation forces on each and the imbalanced neutrino impulses on each.
(Body-body contact over a finite time interval and deformation are both critical in assuring that there is sufficient time for neutrinos to collide so that the expected statistical outcome of the collision can emerge.)
Since there is no net external force applied during the collision, the ``external force'' applied to body 1 can be considered algebraically to be the net neutrino force on body 2 because, without that neutrino force on body 2 to delay its departure, there would be no force transmitted back to body 1.
(Within this paper we elect to avoid the discussing of details concerning the above mentioned deformation because the required complexity of such an analysis would distract from the central points of this section and the effect on momentum transfer would essentially be independent of the details.)
The third law asserts a conservation principle for total body momentum that applies \emph{only} in time intervals \emph{longer} than the fundamental delay times for neutrinos to collide with nucleons.
Since neutrinos are separated from each other on average by \( l_{\nu} \sim 3 \times 10^{-11} \ cm \) but collide with nucleons infrequently because of their very small collision cross sections, the expected distance between neutrinos in the distribution that will collide with a nucleon is \( [n_{{\nu}_{0}} ( \sigma_{\nu,np,cc} + \sigma_{\nu,np,sc,tpf} )]^{-1} \sim 4.3 \times 10^{8} \ cm \) and yields an expected mean time between collisions of \( \sim 1.4 \times 10^{-2} \ sec \).
For an observer in the center-of-momentum frame, there is effectively no net change ever in the total momentum of neutrinos, so the total momentum of the bodies remains \emph{with} the bodies throughout any collision.

This almost immediate action-reaction behavior would be called into question in the cases of ``field'' forces such as with Newtonian gravitation, where the cause-effect delay times must be long considering the typical distances between interacting bodies.
The momentum deficit created by a capture of a neutrino by a nucleon in the first body travels outward from that gravitational body.
The other body senses a gravitational force at a much later time due to more neutrinos impinging on it at that time from the opposite direction.
Total nucleon-neutrino momentum is, however, conserved at all times.

There is no absolute or universal conservation principle for total kinetic energy (and this is expected owing to the artificial construction of \( (1/2) m v^{2} \) to simply relate to work done).
Although total momentum is conserved (according to the third law), that is not sufficient to completely specify \emph{how} momentum is redistributed among bodies and neutrinos.
It is implied that very special conditions (including even nuclear reactions) are required whereby any capture results in almost immediate release of secondary particles to maintain a constant total kinetic energy.
(To the extent that secondary emissions are delayed, strict energy conservation, even when expected to apply, can be violated on even a much larger time scale than the mean time between nucleon-neutrino collisions \( \sim 1.4 \times 10^{-2} \ sec \)).

We note that in all of the above, both gravitational mass and inertial mass are no longer terms that need be employed in explanations of mechanical behavior.
Momentum units seem more appropriate to furnish a basic physical quantity relative to which all other physical quantities are expressed.

\section{Implications of the Neutrino Distribution to Atomic States}

\subsection{The Complete Physically Significant Collision Event}

Since we had no independent criteria with which to estimate individual values for \( n_{{\nu}_{0}} \) or for the mean neutrino energy, we decided to revise downward our previous estimate for the mean energy [from \( 20 MeV \) to \( 2.5 MeV \ ( = 4.0 \times 10^{-6} \ erg) \)].
This then results in a value for \( n_{{\nu}_{0}} \) of \( \sim 3.0 \times 10^{32} \ neutrinos / cm^{3} \) and a related value for the mean distance between neutrinos in the distribution:
\begin{equation} l_{\nu} \ \equiv \ n_{{\nu}_{0}}^{-1/3} \ \sim 1.5 \times 10^{-11} \ cm. \end{equation}

Inclined to believe that the large neutrino energy density would itself create significant disturbance and uncertainty in measurements of particle behavior, we then compared \( h/(2 \pi) \) to the product of this mean distance between neutrinos and the mean neutrino momentum and noted the result showed that the product was actually \( 4 \times 10^{-28} \ erg-sec \) [which is roughly one half \( h/(2 \pi) \)].

We calculated (using equation~\ref{eq:collision_rate}) that there are, on average, \( \sim 300 \) neutrino collisions per second with each nucleon.
This implies that if each separate complete collision event involves only one neutrino, such neutrino collisions with one particular nucleon occur at widely separate intervals of \( \sim (1/300) \ sec, \) further implying that the two neutrinos participating in two separate successive collision events (and traveling at \( c \)) are themselves separated by \( \sim 10^{8} \ cm \ \gg \ l_{\nu} \).
If each complete collision event involves only one neutrino colliding with and being captured by a nucleon every \( (1/300) \ sec \), that nucleon would frequently absorb a substantial momentum impulse equal to that furnished by the neutrino and would thus acquire a kinetic energy several orders of magnitude greater than the binding energy of the inner electrons.
Our calculated large nucleon displacements (or smaller but still very large \emph{nucleus} displacements) would make it impossible to establish stable electron states because there could be no stable pairing of a particular electron with one nucleus.

By far the more likely complete collision event involves at least two or more almost-simultaneous collisions with nearby neutrinos, the first of which puts the nucleon in an excited state that necessarily makes it much more susceptible and favorable to interacting with any nearby neutrino that makes contact (that excited state having a much larger collision cross section).
This complete collision event essentially displaces and then resists the nucleon's motion before it can travel so far as to destroy its bound electron ``orbits''.

To calculate the mean time delay between the first neutrino capture and the next contact with a neutrino (resulting in capture if the entire nucleon is sensitized by that time), we calculated the collision rate (from equation~\ref{eq:collision_rate}) that would result if the collision cross section for a nucleon - neutrino collision is equal to that of the entire (sensitized) nucleon (\( \pi r_{np}^{2} \)).
This yields a collision rate of \( \sim 7 \times 10^{16} collisions / sec \).
The expected time delay between the first neutrino impact in the complete collision event and the second neutrino impact that returns the nucleon momentum to zero is
\[ \sim (1/2) \ (7 \times 10^{16})^{-1} \ \simeq \ 7 \times 10^{-18} \ sec. \]
This time interval is also likely to be much longer than the time it takes for a disturbance from the first neutrino collision to propagate throughout the nucleon and excite it so that it will interact with the next neutrino that makes contact.
[It is possible to consider probable successive neutrino collisions occurring in an even shorter time interval for one nucleus (than for one nucleon alone) if the nucleons that make up that nucleus are so tightly bound to each other as to behave as a single rigid body, but that appears to be a meaningless calculation because the other nucleons, even if they move collectively as a rigid body, would also all have to be simultaneously in an excited state to vastly increase their own collision probability (and interact with their nearest neighboring neutrino).]
Hence, the physically significant collision event would seem to involve the net transfer of one neutrino's momentum to a nucleon for only \( \sim 1 \times 10^{-17} \ sec \), which is approximately the time that it takes for a neutrino to travel \( \sim 2.1 \times 10^{-7} \ cm \), collide with, and offset a large portion of this nucleon's momentum.

During this complete collision event, a nucleus having only one nucleon will move
\[ \sim 1 \times 10^{-17} c / 400 \simeq 7.5 \times 10^{-10} \ cm. \]
Other larger nuclei will move shorter distances during one collision event because of the limited momentum acquired from the first neutrino.
This result (that the magnitudes of these nuclear displacements are small fractions of the atomic diameter) is the main reason (alluded to above) for choosing \( 2.5 MeV \) as the mean neutrino energy.

Since the nucleon moves for the duration of the complete collision event, positively charged nucleons have an opportunity to create electrodynamically a classical electromagnetic disturbance that can affect the nearby electrons.
There are two parts to this disturbance.
A ``static'' or coulombic part results from the transient displacement of the positive nucleon from its equilibrium position, and this transfers momentum to the electrons, which in turn transfer momentum to the body as a whole.
A time varying electromagnetic disturbance is also created due to the momentum signature of the charged nucleon.
This signature is similar to a square wave except that the beginning and end are not perfectly square, indicative of the finite time that it takes to transfer momentum from a neutrino to a nucleon.

A perfect square wave yields a primary frequency whose period is double the length (in time) of the square wave.
In this case, the primary frequency (\( f_{p} \)) would be approximately \( \sim 1 \times 10^{17} / sec \), but this is only a mean value, and there is a statistical spread of such momentum impulse widths.
Square waves also yield higher frequency Fourier components that are related to the primary frequency by the relation
\[ f_{k} = (2k + 1) f_{p}. \]
These have weaker (\( 1/k \)) amplitude ratios relative to the amplitude of the primary frequency.
Signatures that are approximately square but with tapered ends generally suppress the higher frequencies by damping them.
The primary frequency of a particular collision event would generate a dipolar electromagnetic disturbance, and the higher frequency overtones would generate higher multipole electromagnetic disturbances.

It should be noted that all these frequencies are high and have associated wavelengths that are shorter than several tens of Angstroms.
In order that they create classical electromagnetic disturbances and not create photons, the mechanical frequency sources must generate a frequency spectrum that has no sharply defined peaks.

\subsection{Neutrino Momentum Transfer to Nucleons as Part of the Basis of Quantized Electron States}

Approximating the diameter of a nucleon as \( \sim 10^{-13} cm \), then an atomic nucleus with atomic weight of \( 35 \) has a diameter of \( \sim 3 \times 10^{-13} cm \).
Average neutrino spacing is \( \sim 1.5 \times 10^{-11} cm \) and they are moving at speed \( c \) and traveling in random directions.
A \( 2.5 MeV \) neutrino has a mass of \( \sim 1/400 \ AMU \).
Electrons have a mass of \( \sim 1/1862 \ AMU \), and the inner electron states have a diameter of \( \sim 10^{-8} cm \).
This large void between the nucleus and the average radius of the inner most electron states can be explained by the inability to establish a \emph{stable} electron standing wave pattern at distances closer to the nucleus because of the large (and multipole) electromagnetic fluctuations (some being destabilizing to a standing wave) in the near field region of a charged nucleus that is subjected to such small and random displacements hundreds or thousands of times each second.

The inner-most electron states behave as standing waves (closed complete surfaces with non-zero thickness) whose fundamental frequency is determined by the dipolar electromagnetic disturbances produced in the nucleons by the shorter collision events.
These are the most tightly bound electrons and are more sensitive (than are more removed electrons) to this high frequency electromagnetic mode.
An electron further from that nucleus behaving as a loosely-coupled standing wave (also a closed surface) establishes the pattern whose fundamental frequency is matched to one of the lower frequency (associated with longer collision events) dipolar electromagnetic disturbances.
This more remote electron standing wave pattern is more loosely coupled (lower coupling constant) to the motion of the nucleons, and the lower frequency electromagnetic disturbances do not exert much of an influence on the above-mentioned inner electron states.
Since the higher frequency disturbances for the outer electron are too high and too weak to completely upset the fundamental frequency behavior of that electron, these higher frequencies act mainly to superimpose higher multipole perturbations on that standing wave.
These electromagnetic disturbances do not of themselves determine the energies of the electron states that respond to them; they merely are able to transfer energy when necessary to either establish or maintain those states.
The average \emph{positions} of these standing waves relative to the nucleus establish their principle quantum numbers, and the perturbations to the standing waves provide energy differentiation between the states belonging to the same principle quantum number.
Thus, quantization of the electron states results from the quantizing of the electron standing wave patterns (by integral wavelength relationships characteristic of standing waves) caused indirectly by neutrino bombardment, and the nucleon electromagnetic frequencies sustain certain of those candidate electron standing wave patterns but do not of themselves impose quantization.
Other quantization effects that are similarly facilitated by neutrino bombardment are discussed [13, 14].

Because the electromagnetic disturbances that drive the electron states are classical and derive from mechanical motion, the statistical relationships between expectation values of conjugate variables that appear in the Heisenberg Uncertainty Principle apply as well to these electron states, and their numerical products are expected to remain considerably higher than the value of \( h / (2 \pi ) \) (resulting normally in lack of spontaneous photon emission).
As applied to these electron states, the uncertainty relationship is used to describe the fundamental delocalization of the electron mass (and associated charge) that is characteristic of a standing wave (probability amplitude) description even in the absence of external stimuli such as interrogating radiation.

The physical description of a completely free electron differs substantially  from the description of a bound electron.
Because the mass of the electron is so small, it reflects a behavior that is near the boundary between being described as a particle and being described as a wave.
Traveling far from any nucleus, the electron behaves more like a particle.
When it enters the region of a charged nucleus that has an unoccupied electron state, the wave nature predominates in response to the vibrating field of the nucleus, and the electron settles into a stable standing wave pattern.
If the nucleus were not vibrating, there would be nothing to stop the electron from penetrating as a particle to the center and interacting directly with the nucleus.
Thus, these random vibrations of the nucleons caused by the neutrinos and their tendency to behave like targets that are very unfeasible for electrons to impact are what allow for stable atomic structures.

The above description of the neutrino as the coordinator and sustainer of atomic behavior encourages an alternate and more fundamental explanation of black hole behavior.
We believe that the absence of a fully developed omnidirectional neutrino distribution will prevent some stable electron states from persisting in super dense stars.
As soon as the total path-integrated collision cross section along any essentially radial path in such a star is large enough to shield the central core from a large fraction of the incident neutrinos (and their interaction with nucleons as described above), that core coulombically collapses to higher density [15].
This onset condition is specified by
\[ \sigma_{\nu,np,cc} \ n_{np} \ r \ \sim \ 1. \]
Atoms in the outer region similarly coulombically collapse, and the region of collapse continues to rapidly spread outward, releasing large amounts of energy.
In the black hole extreme, there are no sustainable electron states, and the only radiation emission that can emanate from such a black hole comes from impinging external atoms just before they encounter the surface or from purely nuclear phenomena especially transient ones.

\section{Relationship to an Expanding Universe}

We utilize the definition of our universe as being that region of space that is occupied by some material ejected by our big bang, and we consider the possibility that space exists beyond our universe and may contain other universes and other neutrinos from unknown origins.
Current information indicates that our universe is expanding at the rate of \( 50 - 100 km / sec / megaparsec \), but there are differences of opinion as to whether this expansion is accelerating endlessly or decelerating so as lead either to an eventual collapse or to an endless expansion at an ever-decreasing rate.
 
One of the most important issues regarding the neutrino theory of induced gravity concerns the source of those neutrinos.
If the source is some process within the universe that is converting another form of energy to those neutrinos, it must be consistent with the following requirements.
The estimated average mass density of the universe is \( \sim 10^{-29} \ grams / cm^{3} \), and this can be alternately expressed as an average relativistic energy density of \( \sim 3 \times 10^{-9} \ erg / cm^{3} \).
The average neutrino energy density required to account for induced gravity is \( \sim 10^{27} \ erg / cm^{3} \), and most of this neutrino energy continuously escapes from the universe at speeds in excess of other known material and without interacting with any mass particles.
An energy source within the universe capable of sustaining this large neutrino energy density is as-yet undiscovered, and it would have to be distributed over a large region of the universe in order to produce an omnidirectional neutrino distribution.
Nevertheless, a neutrino source within our universe would propel any material at the outer regions of the universe only outward, leading to an infinite expansion for most of the universe material.
 
On the other hand, if the source of the neutrinos is outside our universe (distances \( > 14 \times 10^{9} \) light-years), the neutrinos could be postulated to be part of a primordial environment within which one or many universes were born, or the neutrinos could be postulated to have been the main material ejected from the early stages of many universes.
With any external neutrino source, the ultimate fate of our universe has the same expansion or collapse possibilities as are provided in a Newtonian theory.
However, a neutrino-induced gravity theory with outside neutrino sources, as contrasted with a Newtonian theory, would favor an expansion because the formation of black holes shields significant quantities of mass from neutrinos and thus prevents that mass, once ``hidden'', from having any gravitational influence on other mass bodies. 

\pagebreak

\noindent References and Footnotes

\bigskip
\noindent [1] Giampiero Esposito 1995, Complex General Relativity,
Kluwer Academic Publishers, Dordrecht/Boston/London. \newline
See also G. Esposito et al. 1994 Classical and Quantum Gravity, vol 11, 2939-2950, and 1994 Foundations of Physics Letters, vol 7, 303.

\bigskip
\noindent [2] Smith, L. Sept 13-15, 1997 Proc of International Centennial Symposium on the Electron.  Cambridge University Press (In press)

\bigskip
\noindent [3] \label{fn:order} It is helpful to summarize in advance such first and second-order issues so that when such or higher-order terms are eliminated in approximations, it does not appear arbitrary.

\bigskip
\noindent [4] \label{fn:spherically_symmetric} Our interest here is to continue developing this approach as it applies to spherically symmetric planetary and stellar bodies and to avoid otherwise more complicated mass distributions.

\bigskip
\noindent [5] Gamow, G. 1962 Gravity, Science Study Series, Doubleday, Garden City, N.Y.

\bigskip
\noindent [6] Wheeler, J.A. 1962  Geometrodynamics, Academic Press, New York.
Davis, R.,Jr. 1955 Phys. Rev 97, 766 and 1956 Bull. Am. Phys. Soc, No 4, paper UA5.
See also Brill, D.R. and Wheeler, J.A.  1957 Rev. Mod. Phys 29, 465 and Euwema, R.N. April 1959 ``Neutrinos and their Interaction with Matter'', Ph.D. Thesis, Princeton.
See Lee, T.D., Yang, C.N. 1960 Phys. Rev. Lett 4, 307.

\bigskip
\noindent [7] Information from Internet on Neutrinos:
Bailey, R.T. 1998 The Neutrino Report
Bahcadl and Pinsonneault,1998.
Kim,J. ``An Extraction of the Strong Interaction Coupling Constant'' and Gross-Llewellyn-Smith Sum Rule.
Halpern, L. \& Laurent, B.1964  ``On the Gravitational Radiation of Microscopic Systems'', Nuovo Cimento 23, 728.
Braginskij, V.B. 1966  ``Gravitational Radiation and the Chances for its Experimental Radiation'' Uspekhi Fisiceskikh, Nauk. Vol 86, 436 (1966).
See also Domaschko, M. 1997  ``A Unified Field Theory Free of Attractive Forces''  Internet www.aplg.com/theory.htm.

\bigskip
\noindent [8] Fukuda, Y. Phys. Rev. Lett (in press).
Appollonio, M. 1998 Phys. Lett B. 420, 397.
Schwartzchild, B. 18 Aug 1998 Physics Today.
Hagopian, V. and Baer, H. 1996 Neutrino Mass, Dark Matter, Gravitational Waves, Monopole Condensation, and Light Cone Quantization, Ed. by K. Kursunoglu, Plenum, N.Y.  p.43.  H. B. Prosper Ibid, p. 115.

\bigskip
\noindent [9] a) Einstein, A. 1911 Annalen der Physik 35, 898.
b) Einstein, A. 1905  Annalen der Physik 17.
c) Einstein, A. 1953 The Meaning of Relativity, Princeton University Press, Princeton.
d) Poincare', H. 1952 Science and Method (Dover, New York, 1952).
e) Einstein, A. 1915 Feldgleichungen der Gravitation Pitzber Preuss. Akad. Wiss 844  (Paper in which Einstein introduced cosmological constant).

\bigskip
\noindent [10] Hawking, S. 1992 Phys Rev Lett 69, 406.
Hawking, S. 1976 Phys D 13 191 (1976).
Hawking, S. 1965 Phys Rev Lett 15, 689.
Hawking, S. 1972 Comm. Math. Phys 25, 152.
Hawking, S. 1975 Comm. Math. Phys 43, 199.

\bigskip
\noindent [11] Thorne, K.S. and Zytkow, A.N. 1977 Astrophysical Journal 212, 832.

\bigskip
\noindent [12] Penrose, R. 1965 Phys Rev. Lett 14 51-59; 1976  Gen Rev. Grav. 7, 31; 1982 Proc. Roy. Soc. London A381, 53; 1992 ``Gravity and Quantum Mechanics" in General Relativity and Gravitation.  Proc of Thirteenth Int Conf on General
Relativity and Gravitation, Cordobe, Argentina.
Carter, B. 1971 Phys Rev. Lett 26 331; Hawking, S. and Penrose, R. 1996 The Nature of Space and Time, Princeton University Press, Princeton, p. 105-137.

\bigskip
\noindent [13] \label{fn:electron_states} The result of a small displacement of a positive nucleon relative to the standing wave pattern of an electron sets up a restorative electrical force that behaves in a Hooke's Law fashion.
The nucleon and electron, through the electromagnetic field, form a coupled harmonic oscillator, with the electron motion at least slightly \emph{out of phase} with that of the main time varying components of the driving electromagnetic field disturbances.
This motion, considered alone, would produce a vibration or wobble of the entire system.
It is therefore natural to expect that another closely-related electron state would be available that, if occupied, would reduce the magnitude of this vibration to create the lowest free energy for the system.
However, such a standing wave model for the electron dictates that each such electron state must be distinguishable from all others (relative to precise shape so as to not substantially overlap) because otherwise the two similar standing waves will quickly destructively interfere.
Two such similar standing waves might coexist briefly if their phase difference has just the right value, but that fortuitous correlation can only be transient because there is no physical tendency of the system to preserve an exact phase difference.
This neutrino-modified coupled harmonic oscillator atomic structure viewpoint seems a viable model that can explain the recently discovered reaction or response of the nucleus to a double shell excitation of valence electrons.

\bigskip
\noindent [14] \label{fn:pauli_exclusion} There is another important aspect to the randomness (randomness in time between successive initiating neutrino collisions and randomness in direction of impact) of the impulses delivered to the nucleons of the nucleus by the neutrinos.
If one accepts that electrons establish modes of vibration each of which is driven by a particular electromagnetic frequency emanating from the neutrino-perturbed/induced motions of the charged nucleons, those separate nucleon motions must not be so highly organized as to tightly couple over extended time intervals with the resistanceless harmonic motions associated with one or more of those electron modes.
If any substantial long term lock-on were to occur, large amplitude motion of the electron would likely result due to the ability of the nucleus to transfer large amounts of energy while the driving force on the electron is in phase with its velocity.
This is extremely unlikely to occur, however, because neutrinos in general are \emph{not} correlated as to spacing or direction of travel.
Compared to the extremely short duration of a complete collision event, the mean time between successive collision events involving one nucleus is very long and unlikely to sustain any correlations that might have been established \emph{during} the previous collision event.

\bigskip
\noindent [15] \label{fn:fusion_collapse} The transition from light atomic elements throughout to dense compacted nucleons in such stars probably occurs in several stages, each marked initially by the collapse of the electron states of some lighter elements and the freeing of their nucleons and electrons.
Other atoms can, in a fusion process, capture those freed particles to form heavier elements which have a higher sustainability because of their increased probability to collide with the decreased neutrino population.
This will continue until no heavier elements can be formed, at which point the final collapse is to a black hole state where there are no longer any large voids separating any nucleons from each other.

\end{document}